\documentclass[]{aa}  

\usepackage{graphicx}
\usepackage{txfonts}
\usepackage{epsfig}
\usepackage{times}
\usepackage{rotating}
\usepackage{setspace}
\usepackage{natbib}
\def\gtsim{\mathrel{\spose{\lower.5ex \hbox{$\mathchar"218$}}
     \raise.4ex\hbox{$\mathchar"13E$}}}
\def\ltsim{\mathrel{\spose{\lower.5ex\hbox{$\mathchar"218$}}
     \raise.4ex\hbox{$\mathchar"13C$}}}

\def\gtsim{\mathrel{\spose{\lower.5ex \hbox{$\mathchar"218$}}
     \raise.4ex\hbox{$\mathchar"13E$}}}
\def\ltsim{\mathrel{\spose{\lower.5ex\hbox{$\mathchar"218$}}
     \raise.4ex\hbox{$\mathchar"13C$}}}

\def\Ha{${\rm H}_{\alpha}$}
\def\Hb{${\rm H}_{\beta}$}
\def\Hg{${\rm H}_{\gamma}$}
\def\Hd{${\rm H}_{\delta}$}

\def\Fe{\rm Fe}

\def\ZH{[$Z/{\rm H}$]}
\def\MH{[$M/{\rm H}$]}

\def\kms{$\rm km\;s^{-1}$}

\def\spose#1{\hbox to 0pt{#1\hss}}

\def\kms{$\rm km\;s^{-1}$}

\def\Hb{${\rm H}{\small{\beta}}$}
\def\Ha{${\rm H}{\small{\alpha}}$}

\begin{document}

   \title{Stellar population properties for a sample of hard X-ray AGNs}


   \author{L. Morelli
          \inst{1,2}
          \and
          V. Calvi\inst{1,3}
        \and
        N. Masetti\inst{4}
        \and
        P. Parisi\inst{5}
        \and
        R. Landi\inst{4}
        \and
        E. Maiorano\inst{4}
        \and
        D. Minniti\inst{6,7,8}
        \and
        G. Galaz\inst{6}
  }

   \institute{Dipartimento di  Fisica e Astronomia ``G. Galilei", Universit\`a di Padova,
              vicolo dell'Osservatorio 3, I-35122 Padova, Italy\\
              \email{lorenzo.morelli@unipd.it}
              \and
             INAF-Osservatorio Astronomico di Padova,
              vicolo dell'Osservatorio~5, I-35122 Padova, Italy
              \and
              Space Telescope Science Institute, 3700 San Martin Drive,
              Baltimore, MD 21218, USA
              \and
              INAF – Istituto di Astrofisica Spaziale e Fisica Cosmica di Bologna, 
              via Gobetti 101, 40129 Bologna, Italy
              \and
              INAF -- Istituto di Astrofisica e Planetologia Spaziali, Via del Fosso del Cavaliere 100, 
              Roma I-00133, Italy
              \and
               Departamento de Astronom\'ia y Astrof\'isica, Pontificia Universidad Cat\'olica de Chile, 
               Casilla 306, Santiago 22, Chile
               \and
               Vatican Observatory, V00120 Vatican City State, Italy
               \and
               Departamento de Ciencia Fisicas, Universidad Andres Bello, Santiago, Chile}

   \date{\today}

 
  \abstract
   {}
{The aim of this paper is to study the stellar population of galaxies
  hosting an active galactic nucleus (AGN). We studied a sub-sample of
  hard X-ray emitting AGNs from the {\it INTEGRAL} and {\it Swift}
  catalogs which were previously identified and characterized through
  optical spectroscopy. Our analysis provides complementary
  information, namely age and metallicity, which is necessary to
  complete the panoramic view of such interesting objects.}
{We selected hard X-ray emitting objects identified as AGNs by 
  checking their optical spectra in search for absorption lines suitable 
  for the stellar population analysis. We obtained a final sample consisting of 20 
  objects with redshift lower than 0.3.  We used the full-spectrum 
  fitting method and, in particular, the penalized pixel one applying the 
  PPXF code. After masking all the
  regions affected by emission lines, we fitted the spectra with the
  MILES single stellar population templates and we derived
  mass-weighted ages and metallicities. }
{Most of the objects in our sample
show an old stellar population, but three of them are characterized by
a bimodal distribution with a non negligible contribution from young
stars. The values of the mass-weighted metallicity span a large range
of metallicity with most of them slightly above the solar value.  No
relations between the stellar population properties and the
morphological ones have been found. }
{}

   \maketitle
%

\section{Introduction}

The massive work carried out by \cite{masetti2004, masetti2006c,
  masetti2006a, masetti2006d,
  masetti2006b,masetti2008,masetti2009,masetti2010,masetti2012},
\cite{landi2007}, \cite{parisi2009,parisi2012}, and
\cite{maiorano2011} (hereafter, Papers I-XIII) supplied the
astronomical community with a catalog\footnote{The up-to-date version
  of this catalog is available on
  http://www.iasfbo.inaf.it/$\sim$masetti/IGR/main.html} of hard X-ray
sources for which the identification and the main physical parameters,
computed using the multiwavelength information available in the
literature, are provided. In total, more than 250 objects in the
20-200 keV range, observed during the {\it INTEGRAL}
\citep{Winkler2003} and {\it Swift} \citep{gehrels2004} missions,
$\sim$200 of them using the IBIS instrument \citep{ubertini2003} and
$\sim$60 using the BAT one \citep{barthelmy2004}, were studied and
classified.  In particular, to unveil the nature of most of these
objects, an optical follow-up was mandatory since only optical spectra
permit an accurate source classification and provide fundamental
parameters.  In Papers I-XIII we measured the flux of the most
important emission lines existing in the optical part of the spectrum
with the main aim of identifying the object but also of investigating
the properties of the host galaxies as, for example, the Compton
nature of these objects and an estimate of the mass of the central
black hole in broad emission line AGNs.  Nevertheless, information
regarding the stellar population of the galaxy hosting the AGNs is
missing in the catalog. Therefore, stellar populations are new,
important and complementary pieces of information which should be
included in the catalog and which shed light on the properties of the
AGNs and their host galaxies.

In the last decade several authors focused their attention on the
relation between the nuclear activity and the star formation rate
\citep{ivanov00,gonzdelgado01,joguet01,ho03,lamura09,draper11,cracco11,vaona12}.
Such relation can give important clues about the fate of the gas
fuelling the central black hole \citep{hopkins06} and its influence on
the central part of the host galaxy \citep{sarzi05}.

According to simulations \citep{dimatteo2005,springel05,fontanot11}, a
merging episode involving galaxies rich in gas induces radial gas
inflows which feed the black hole and, consequently, enhance the
central star formation. Then, the AGN feedback wipes out basically all
the remaining gas and dust \citep{hopkins06,rigopoulou09}, halting the
star formation. However, in the last years the proposed scenarios
became even more complicated. In fact, the model by \cite{novak11}
showed that the AGN is a cyclic process, and to account for the
observational properties of this class of objects it is necessary to
consider different processes acting on different scales, like mechanic
feedback inside few pc from the nucleus, radiative feedback and
consequent cooling flow of gas on a scale of few kpc, and SN winds,
which are relevant on scales of tens of kpc \citep{ciotti10}.

Some observational works \citep{tremonti2007,feruglio10,mckernan2010} 
stated that AGNs could have played a relevant role in halting the star 
formation in massive host galaxies. \cite{schawinski2007} suggested that, 
if the accretion of material onto the central black hole is powerful 
enough, then the feedback could stop any star formation activity in the 
nucleus of the hosting galaxy and even beyond. As a consequence of this 
process the stellar populations in the AGN host galaxies is old and this, as 
suggested by \citet{schawinski2007} and \citet{faber07}, causes the 
host galaxies to move from the blue cloud to the red sequence of the 
galaxy optical color diagram. Moreover, \cite{bluck2011} computed the 
average energy output per galaxy due to AGN showing that this is at least 
35 times larger than the binding energy of a typical massive galaxy.

Since an invaluable piece of information to understand the processes
of formation and evolution of galaxies is imprinted in their stellar
populations, in the recent years some studies aimed to investigate the
properties of the stellar populations of the host galaxy have been
conducted. Studying the stellar populations in the nuclear region and
their radial profile of a sample of AGNs, mostly Seyfert 2 (hereafter
Sy2), it was found that about one third of the galaxies in the sample
shows an old bulge-like stellar population in the center
\citep{ho03,cidfernandes2004,chen09} and that the number of objects
with old stellar population increases to about two thirds when the
outer regions of low luminosity AGN hosts are investigated
\citep{cidfernandes2004}.

In a recent paper \cite{lamura12} investigated the connection between
stellar population and mass in a large sample of type 1 and 2
AGNs. They found that the mass of the stellar component is a key
ingredient to study the star formation history \citep{mannucci10} of
galaxies and, taking it into account, they suggested an evolutionary
sequence moving from starburst galaxies to AGNs
\citep{davies07,schawinski2007}.

Essentially, the observations show a variety of results even when they
are restricted to the analysis of only Sy2 galaxies for which broad
lines are weak or absent \citep{lawrence1987}. This could be due to
the difficulties in choosing homogeneous samples of objects as well as
to an intrinsic complexity in interpreting the results, especially
considering that the AGN theoretical model is still debated.

In this scenario, our results will be useful to understand the
properties of AGN hosts and the consequences of the AGN feedback on
them, as well as to help in testing the predictions of theoretical
models.

The aim of this paper is, therefore, to present the stellar
populations properties for a sample of AGN hosting galaxies which were
identified as the optical counterparts of hard X-ray emitting sources
in Papers I-XIII.

In Section 2 we described the sample selection and the characteristics
of the data.  In Section 3 we illustrated the tools we used and the
procedure we followed to obtain the final results.  In Section 4 we
presented and discussed our results concerning the stellar
populations, in particular the age and the metallicity derived from
the stellar population fitting procedure. Finally, in Section 5 we
summarized results and conclusions.


\section{Sample Selection and Data Properties}
\subsection{The Sample}

The aim of this paper is to complete the catalog of optical
counterparts of hard X-ray sources detected in the massive survey
performed by our collaboration (papers I-XIII), for which we have
already detected and measured the emission line features, with the
study of the stellar population properties of the hosting galaxy.
Consequently, for a subsample of objects in the catalog complete
information in the X-ray and optical range of the spectrum will be
available.  In order to choose the galaxy sample we started selecting
all those objects identified as AGNs in Papers I-XIII.  In
  detail, these hard X-ray sources were selected among those with
  unidentified nature belonging to the {\it INTEGRAL} and {\it
      Swift} surveys (e.g., \citealt{Bird2010,cusumano2010}). The
  selection method consisted in choosing sources containing a single
  soft X--ray object within their arcmin-sized hard X--ray error 
    circle. According to \citet{Stephen2006} this is, with a very high
    degree of probability, the lower-frequency counterpart of the
    high-energy emission. Given the arcsecond precision with which
    the position of soft X-ray sources is available, this technique
    reduces the sky area for the search of the optical counterpart by
    a factor $>$10$^3$, easily allowing to pinpoint the actual
    optical counterpart on which we could, eventually, perform optical
    spectroscopy to determine its nature.  Our survey detected 158
  AGNs of different type, but since the study of the stellar
  populations was not the principal aim of the investigation when the
  spectra were acquired, many of them are characterized by a short
  exposure time which does not guarantee either a quality nor a
  signal-to-noise ratio ($S/N$) suitable for performing a reliable
  analysis of their stellar population properties.  Out of the initial
  sample of optical counterparts of AGNs, for the following analysis
  we considered only the extracted spectra with $S/N \ge 20$ and for
  which the important absorption lines (i.e. \Hb, {\rm Mg}, and {\rm
    Fe}) were not strongly contaminated by broad-band emission lines
  or residual sky subtraction.
Therefore, the final sample comprised 20 objects, mostly
classified as Sy2, with soft and hard X-ray emission detected, and
with gas emission lines measured.
All these galaxies lie in the redshift range $0.008\le {\rm z}\le
0.3$, and their properties are listed in Table \ref{Tab:sample}.

\begin{table*}[t]
\caption{Properties of the galaxies in the sample.}
\label{Tab:sample}
\begin{center}
\begin{tabular}{l l c c c c c l}
\hline
\hline
\noalign{\smallskip}
\multicolumn{1}{c}{Galaxy} &
\multicolumn{1}{c}{Galaxy} &
\multicolumn{1}{c}{Type} & 
\multicolumn{1}{c}{$T$} & 
\multicolumn{1}{c}{Class} &
\multicolumn{1}{c}{$z$} &
\multicolumn{1}{c}{$B_{\rm T}$} & 
\multicolumn{1}{c}{Source} \\
\noalign{\smallskip}
\multicolumn{1}{c}{name} & 
\multicolumn{1}{c}{alt. name} & 
\multicolumn{1}{c}{}&
\multicolumn{1}{c}{}&
\multicolumn{1}{c}{}&
\multicolumn{1}{c}{}&
\multicolumn{1}{c}{(mag)} &
\multicolumn{1}{c}{} \\
\noalign{\smallskip}
\multicolumn{1}{c}{(1)} &
\multicolumn{1}{c}{(2)} &
\multicolumn{1}{c}{(3)} &
\multicolumn{1}{c}{(4)} &
\multicolumn{1}{c}{(5)} &
\multicolumn{1}{c}{(6)} &
\multicolumn{1}{c}{(7)} &
\multicolumn{1}{c}{(8)} \\
\noalign{\smallskip}
\hline	
\noalign{\smallskip}	    
IGR J01528-0326      &PGC 6966                &SA(s)c?  &   5.0    &likely Sy2 & 0.017 &14.11&  \citet{masetti2008}   \\
IGR J02524-0829      &LEDA 10875              &Sa?      &   1.7    &Sy2        & 0.017 &15.03&  \citet{masetti2009}   \\
IGR J04451-0445*     &LEDA 1053623            &S?       &   1.7    &likely Sy2 & 0.076 &17.10&  \citet{masetti2010}   \\
IGR J18244-5622      &IC 4709                 &Sa  	&   1.5    &Sy2        & 0.017 &14.42&  \citet{masetti2006d}   \\
IGR J18308+0928      &LEDA 1365707            &E?       &  -2.0    &Sy2        & 0.019 &15.06&  \citet{masetti2010}   \\
PBC J0041.6+2534     &NGC 0214                &SAB(r)c  &   5.0    &Sy2/LINER  & 0.015 &12.94&  \citet{parisi2012}   \\
PBC J0759.9+2324     &MCG+04-19-017           &Sab      &   2.2    &Sy2        & 0.029 &14.80&  \citet{parisi2012}   \\
PBC J0919.9+3712     &IC 2461                 &Sb 	&   3.3    &Sy2        & 0.008 &14.63&  \citet{parisi2012}  \\
PBC J0954.8+3724     &IC 2515                 &Sb 	&   3.0    &Sy2        & 0.019 &15.02&  \citet{parisi2012}   \\
PBC J1246.9+5432     &LEDA 43101              &Sa 	&   1.0    &Sy2        & 0.017 &13.60&  \citet{parisi2012}   \\
PBC J1335.8+0301     &NGC 5231                &SBa 	&   1.0    &Sy2        & 0.022 &14.29&  \citet{parisi2012}   \\
PBC J1344.2+1934     &PGC 048674              &E?       &  -1.9    &Sy2/LINER  & 0.027 &15.58&  \citet{parisi2012}    \\
PBC J1345.4+4141     &NGC 5290                &Sbc?     &   4.0    &Sy1.9      & 0.009 &13.30&  \citet{parisi2012}   \\
PBC J1546.5+6931     &PGC 2730634             &S?	&   0.5    &Sy1.9      & 0.037 &16.08&  \citet{parisi2012}   \\
Swift J0134.1-3625   &LEDA 5827               &SA0      &  -1.2    &Sy2        & 0.029 &14.03&  \citet{parisi2009}   \\
Swift J0501.9-3239   &LEDA 17103              &SB0/a?(s)&   0.1    &Sy2        & 0.013 &13.87&  \citet{parisi2009}   \\
Swift J0601.9-8636   &LEDA 18394              &Sb? 	&   2.8    &Sy2        & 0.006 &13.51&  \citet{landi2007}   \\
Swift J0811.5+0937   &USNO-A2.0 0975-05763590 &- 	&   -      &XBONG      & 0.286 &  -  &  \citet{parisi2009}    \\
Swift J0911.2+4533   &LEDA 2265450            &S?  	&   3.3    &Sy2        & 0.027 &16.47&  \citet{parisi2009}   \\
Swift J1238.9-2720   &ESO506–G027             &S0       &  -0.8    &Sy2        & 0.024 &14.66&  \cite{landi2007}   \\
\noalign{\smallskip}
\hline				    	    			 
\end{tabular}
\end{center}
\begin{minipage}{18.5cm}
\begin{small}
NOTES: Col. (1): name of the galaxy. Col. (2): alternative name of the
galaxy. Col. (3): morphological type derived from the Third Reference
Catalogue of Bright Galaxies \citep{RC3} and from the Hyperleda
database \citep{hyperleda}. Col. (4): numerical morphological
  type $T$ from the Hyperleda database. Col. (5): classification of
the AGN.  Col. (6): redshift of the object. Col. (7): total apparent
  magnitude in the B-band derived from the Hyperleda database. Col. (8):
reference source for the galaxy. * The identification of this object
is still not confirmed.

\end{small}
\end{minipage}
\end{table*}

   \begin{figure}
              \includegraphics[angle=0.0,width=0.491\textwidth]{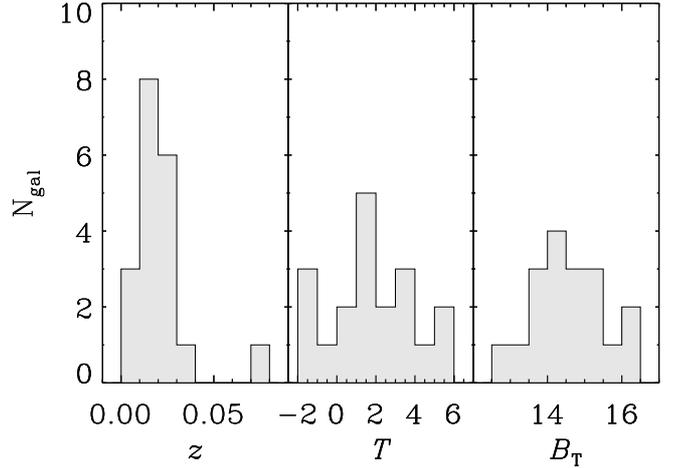}
    \caption{ Histograms showing the distribution of the main
      properties of the galaxies in our sample, i.e. from left to
      right redshift {\it z}, morphological type {\it T} and total magnitude in
      the B-band {\it B}$_T$.}
    \label{Fig:esempio}
   \end{figure}

%
%
\subsection{The data}
The spectroscopic observations of the sample galaxies were carried out
with a variety of setup using  different telescopes. In detail:
\begin{itemize}
\item the 3.58m ``Telescopio Nazionale Galileo'' (TNG) in La Palma, Spain;
\item the 2.1m telescope of the Observatorio Astronomico
Nacional in San Pedro Martir, Mexico;
\item the 1.5m at the Cerro Tololo Interamerican Observatory
(CTIO), Chile;
\item the 1.52m ``Giandomenico Cassini'' telescope of the Astronomical 
Observatory of Bologna, in Loiano, Italy.
\end{itemize}
Some spectra were also retrieved from the Sloan Digital Sky Survey
(SDSS) archive \citep{adelman2006,adelman2008} as well as from the
Six-degree Field Galaxy Survey (6dFGS) archive \citep{jones2004}.

We refer to  Papers I-XIII for detailed explanations on the
observing setup, data reduction, calibration and previous analysis. In
column 6 of Table \ref{Tab:sample} the reference paper for each
  object is listed.  In this paper we just summarized the basic
useful properties of the spectra used in the following analysis.

The sample consists of 20 galaxies spectra acquired in long slit mode.
The wavelength range between 3800 and 7500 \AA\ was covered with a
reciprocal dispersion between $\sim0.8$ and $\sim5.7$
\AA\ pixel$^{-1}$ after pixel binning. This corresponds to an
instrumental velocity dispersion within the range $61$
\kms$\lesssim\sigma_{\rm inst}\lesssim300$ \kms\ at 5500 \AA.
%

\section{Measurements of the Stellar Populations}
\label{sec:misure}

The stellar population properties, namely age and metallicity, were
measured mainly from the following absorption features: \Ha\/ line
($\lambda$ 6563\AA), \Hb\/ line ($\lambda$ 4861\AA), \Hg\/ line
($\lambda$ 4340\AA), \Hd\/ line ($\lambda$ 4102\AA), Mg~{\small I}
line triplet ($\lambda\lambda$ 5164, 5173, 5184\AA), and \Fe\/ lines
($\lambda\lambda$ 5270, 5335\AA). With few exceptions, it was not
possible to obtain information either from the blue part ($\lesssim
3800$\AA) of the spectrum, because of the low efficiency of the
optics, nor from the red part ($\gtrsim 7500$\AA), due to the
residuals of the strong emission lines of the sky in this region.

As done by \citet{onodera12} we applied the penalized pixel fitting
(pPXF; \citealt{capems04}\footnote{Program available on
  http://www-astro.physics.ox.ac.uk/~mxc/idl/}), including the linear
regularization of the weights \citep{presetal92}, to derive the
  distribution of the mass fraction in different age and metallicity
  bins, and the Gas AND Absorption Line Fitting
\citep[GANDALF,][]{sarzetal06} IDL\footnote{Interactive Data Language
  is distributed by Exelis Visual Information Solutions.} packages,
adjusted for dealing with the sample spectra.

Even if in Sy2 galaxies the featureless continuum and broad lines are
weak or absent \citep{lawrence1987}, in the fitting procedure we
allowed the code to use also broad-band components. To account
  for the effect of dust and possible residuals of the data reduction
  procedure, we adopted a low order multiplicative polynomial in the
  template fitting. This has the advantage to make our method more
  sensitive to the absorption lines than to the continuum shape and,
  therefore, less sensible to the effects of reddening.  However,
  we could not completely rule out the possibility of underestimating
  the weight of a very reddened young stellar components in deriving
  the composite stellar populations.

  For each spectrum, we fitted a linear combination of 156 template
  stellar spectra \citep{vazdekis2010} from MILES
  \citep{sancetal06lib} library (${\rm FWHM} = 2.54$ \AA\ spectral
  resolution \citealt{beifetal11}) to the observed galaxy spectrum by
  performing a $\chi^2$ minimization in pixel space. Since the
    resolution of the sample spectra spans a large range of values
    (always lower than the template stellar spectra one), for each galaxy
    it was necessary to convolve each template with the line-of-sight
    velocity distribution (LOSVD) and to rebin both the template and
    the galaxy spectrum to match their dispersions before running the
    fitting code.  

We adopted the Salpeter initial mass function \citep{salp55}, 26 ages
ranging from 1 Gyr to 17 Gyrs, and 6 metallicities \MH\/ from -1.71 to
0.22.  Simultaneously, we fitted the observed spectra using emission
lines in addition to the stellar templates.

The extracted spectra on which we performed the measurement were the
same used to derive the emission lines properties in Papers
I-XIII. This guarantees to perform the analysis in the same spatial
regions used for the detection and measurements of the emission
lines. The linear scale on which we were measuring galaxy properties
depends on the combination of redshift and slit aperture and it ranges from
200 pc to 2 kpc. Swift J0811.5+0937 is the only exception, having a
linear scale of 6 kpc. On the basis of above, in almost all the cases
the region we were investigating was big enough to allow a study of
the stellar populations of the bulge central region surrounding the
AGN.

The example of fitting procedure shown in Figure \ref{Fig:esempio}
(left panel) refers to galaxy PBC~J0954.8+3724 and proves the good
quality of the fit we obtained for all the galaxies. From the fitted
single stellar population (SSP) we, then, derived the stellar mass
fraction within each age and metallicity interval (right panel).
   \begin{figure*}
              \includegraphics[angle=0.0,width=0.991\textwidth]{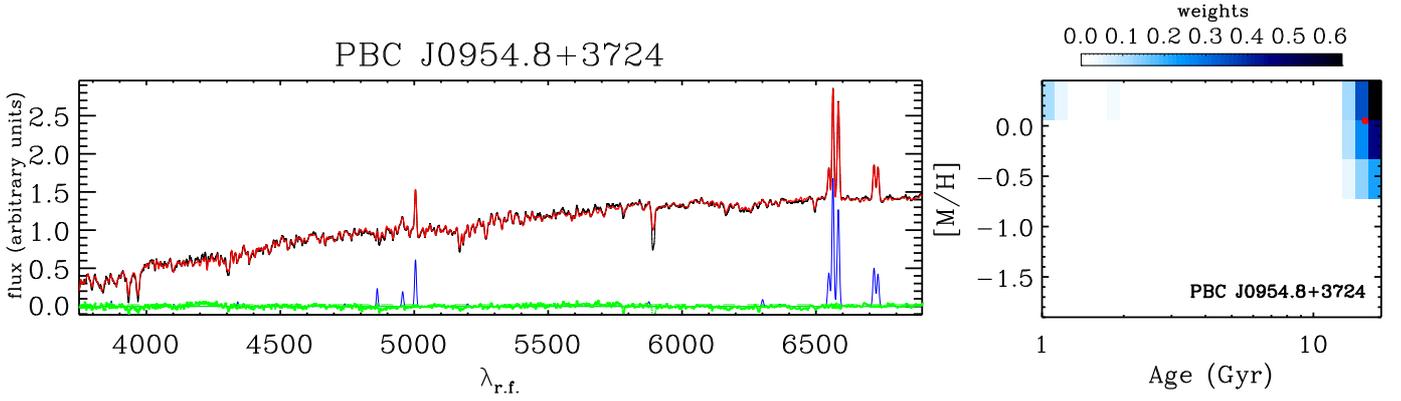}
    \caption{Left panel: the galaxy spectrum (black line) is compared
      to the fitted linear combination of template stellar spectra
      (red line) and emission lines (blue line). The green dots show
      the residuals.  Right panel: age and metallicity obtained from
      the spectral fitting. The color scale refers to the mass
      fraction in each bin of age and metallicity. The red dot represents the
             mass-weighted age $\langle t/Gyr \rangle_M$ and the mass-
             weighted metallicity $\langle\,$\MH$\,\rangle_M$.}
    \label{Fig:esempio}
   \end{figure*}
%
%
\section{Results and Discussion}
\label{sec:risultati}

We applied the procedure described in Section \ref{sec:misure} to
all the galaxies in our sample. 

\begin{figure*} 
              \centering
              \includegraphics[angle=0.0,width=0.450\textwidth]{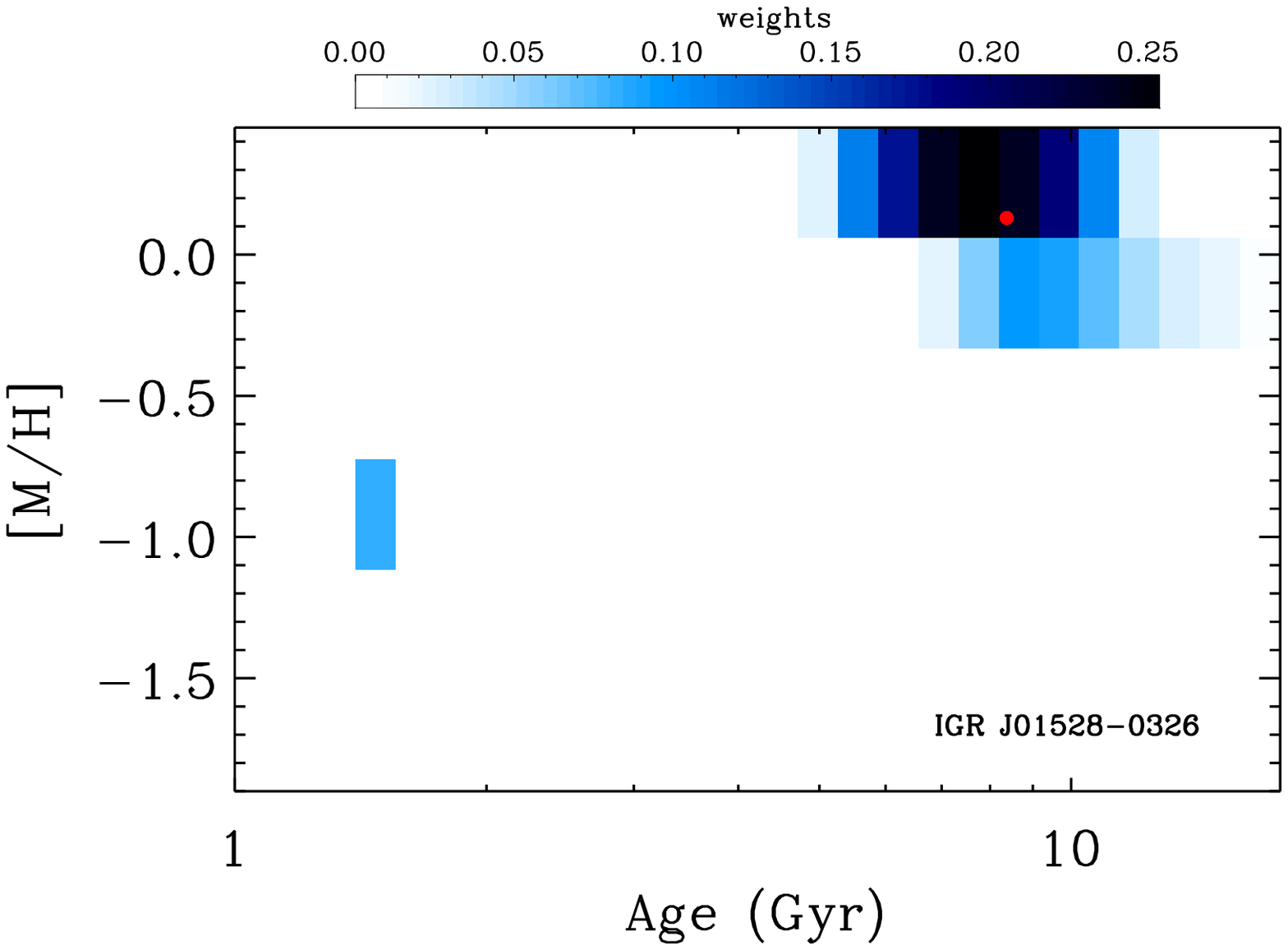}
              \includegraphics[angle=0.0,width=0.450\textwidth]{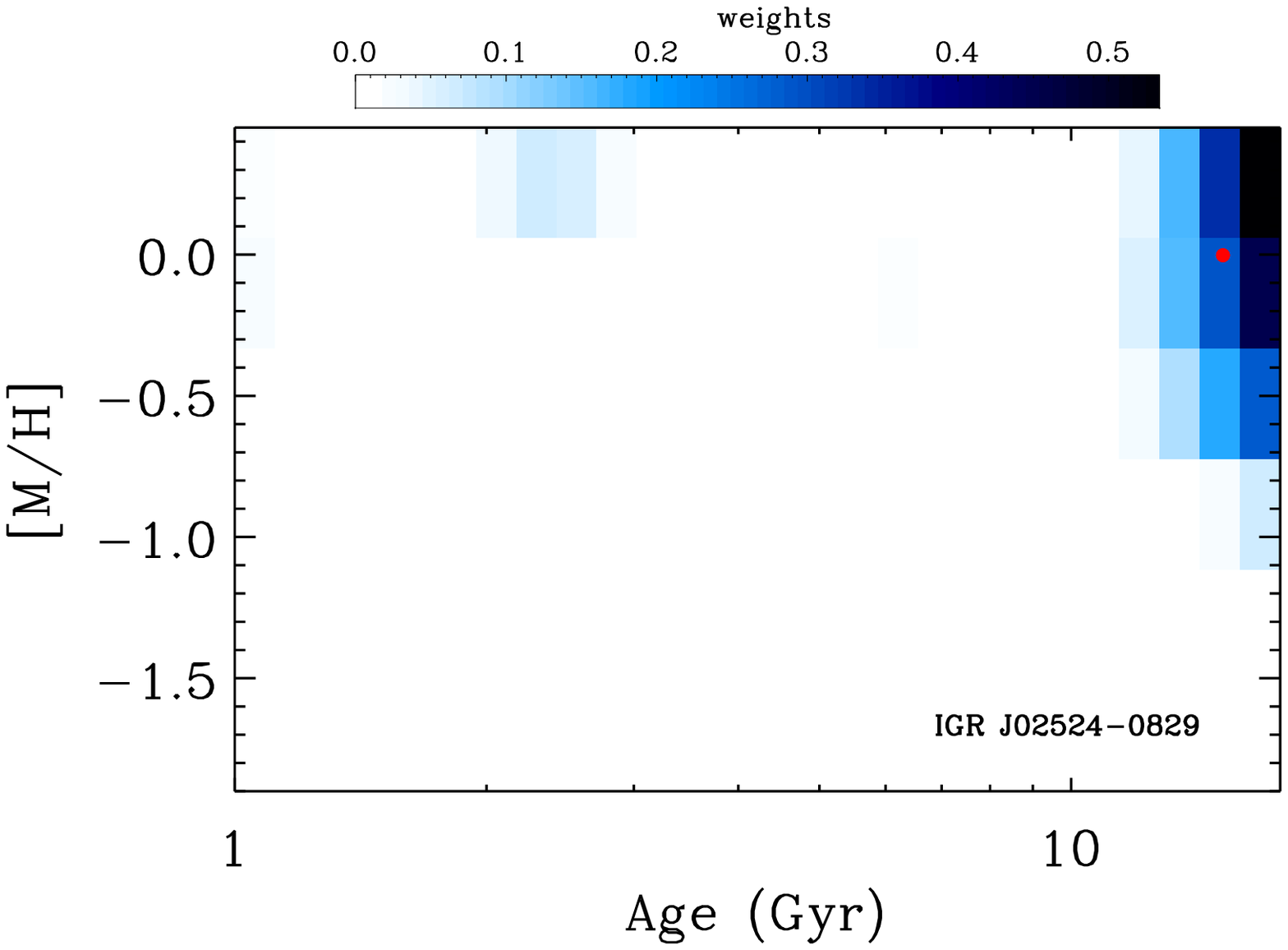}
              \centering
              \includegraphics[angle=0.0,width=0.450\textwidth]{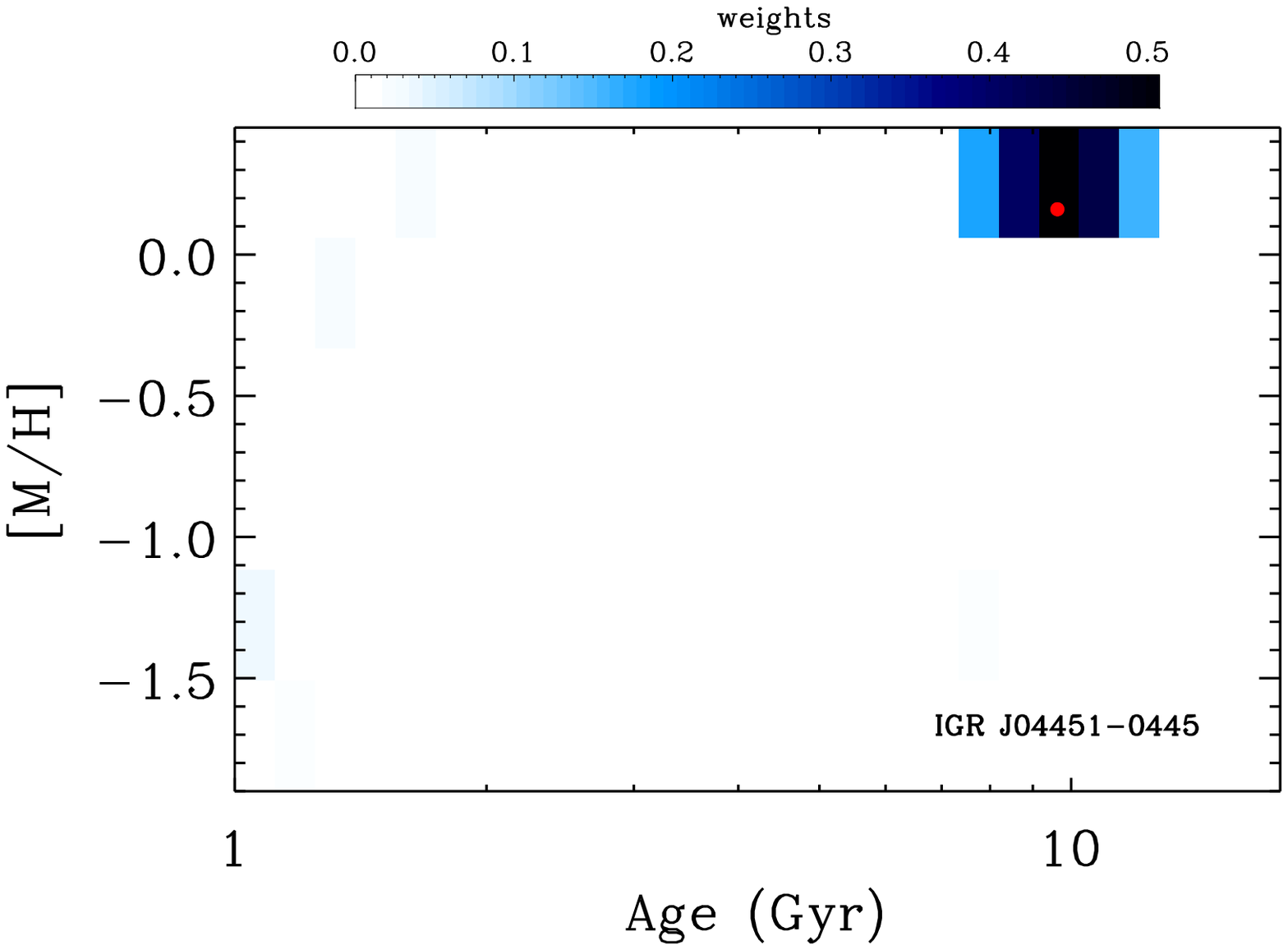}
               \includegraphics[angle=0.0,width=0.450\textwidth]{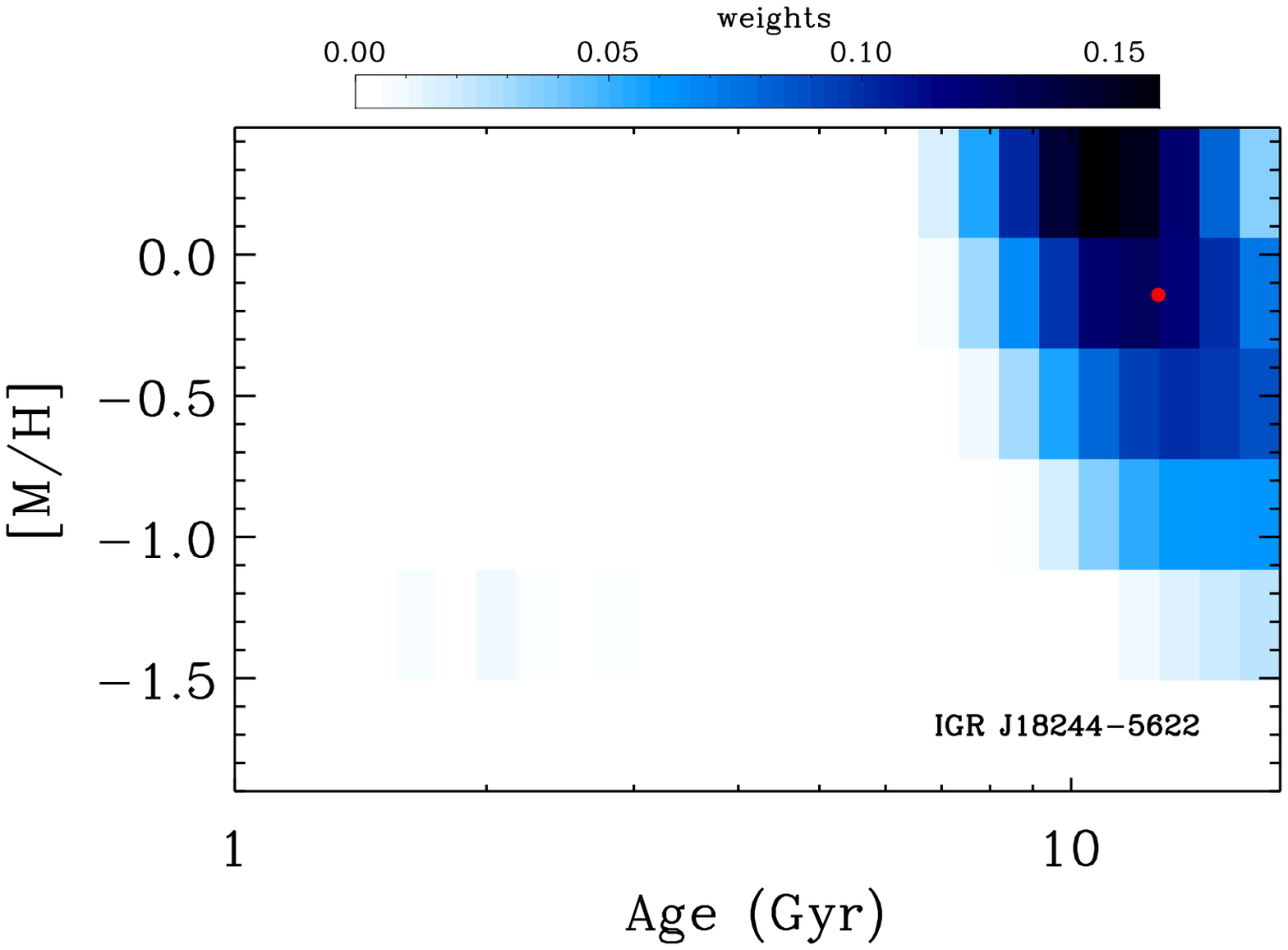}
              \centering
              \includegraphics[angle=0.0,width=0.450\textwidth]{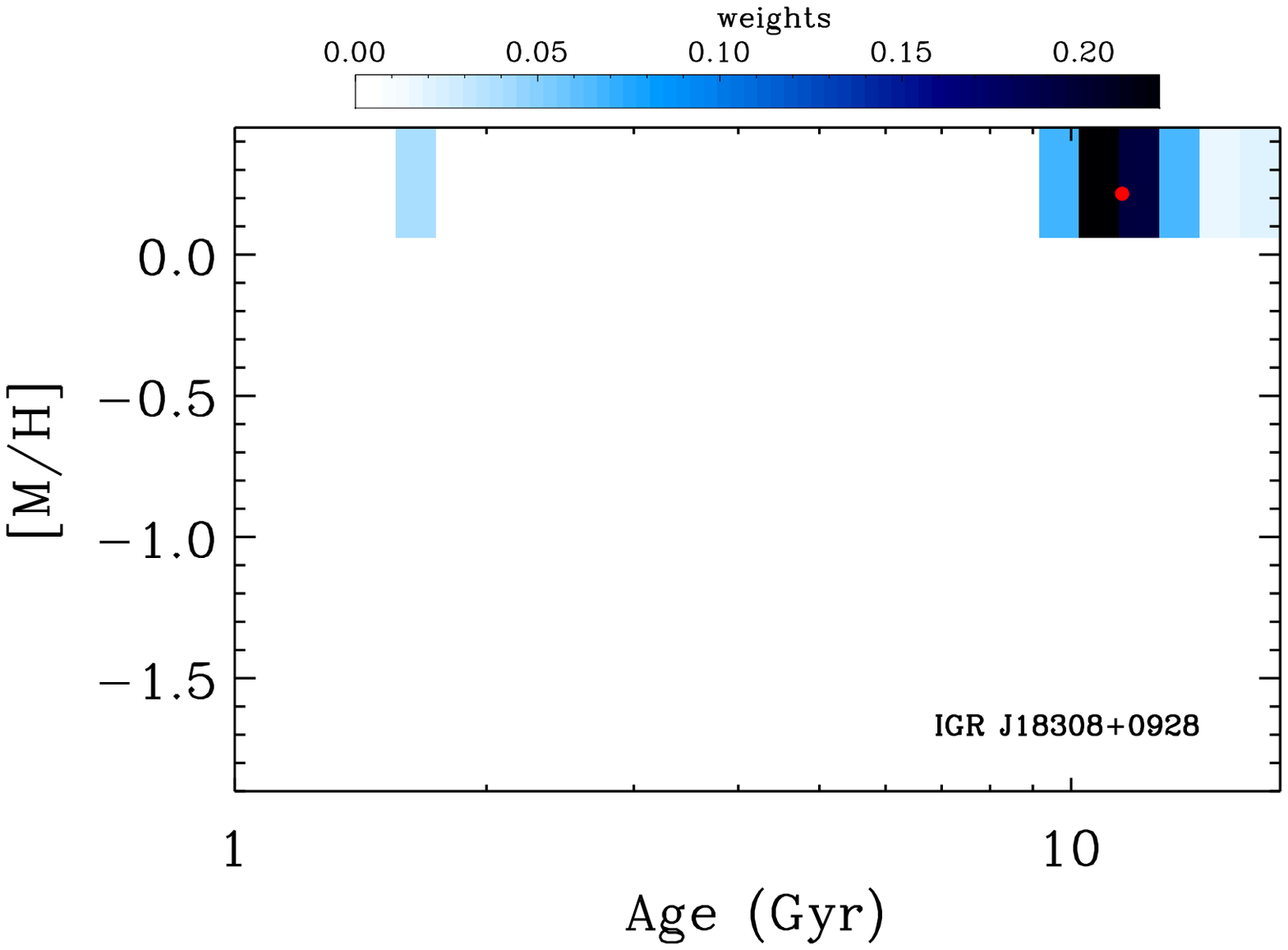}
              \includegraphics[angle=0.0,width=0.450\textwidth]{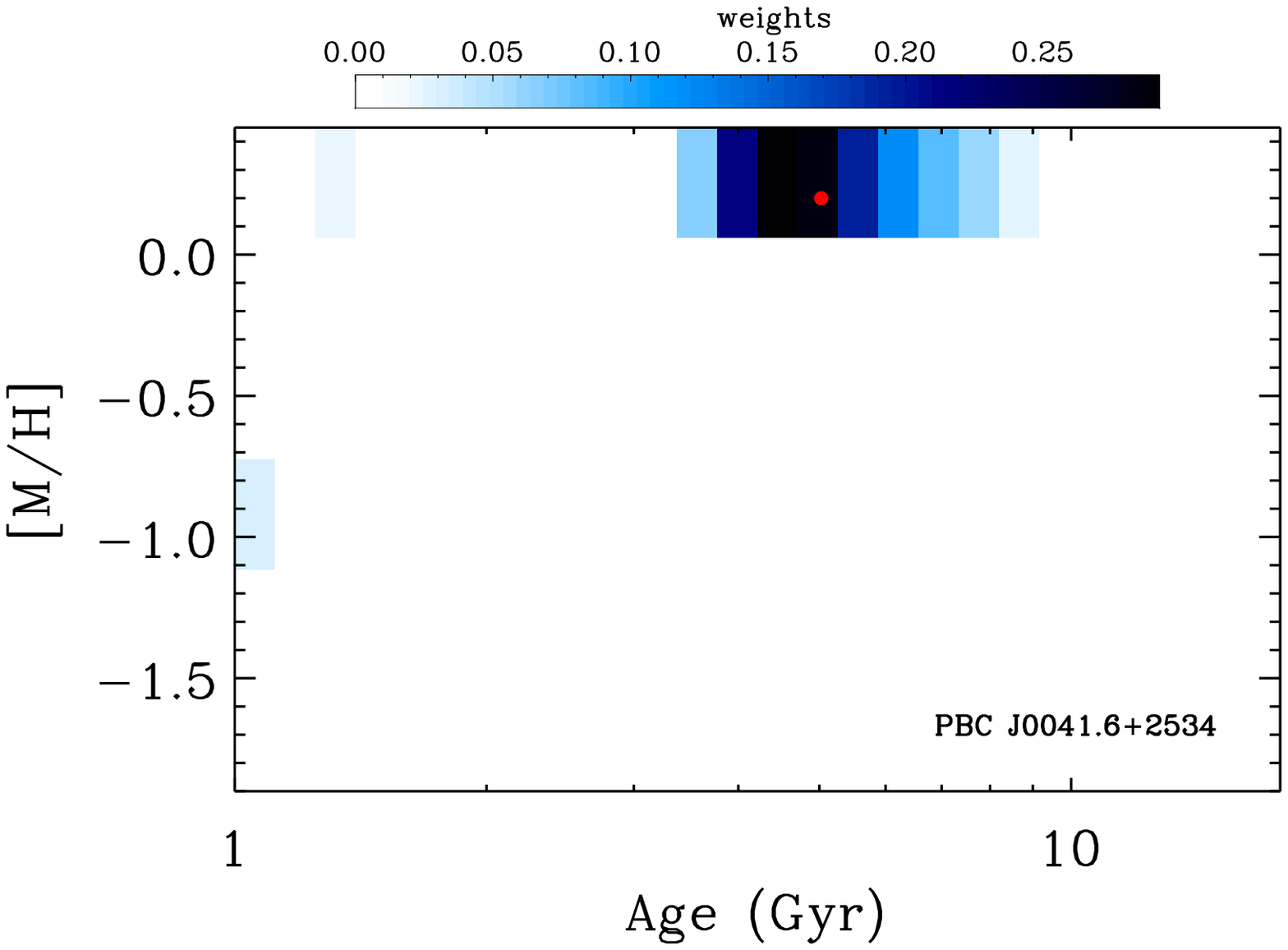}
              \centering
              \includegraphics[angle=0.0,width=0.450\textwidth]{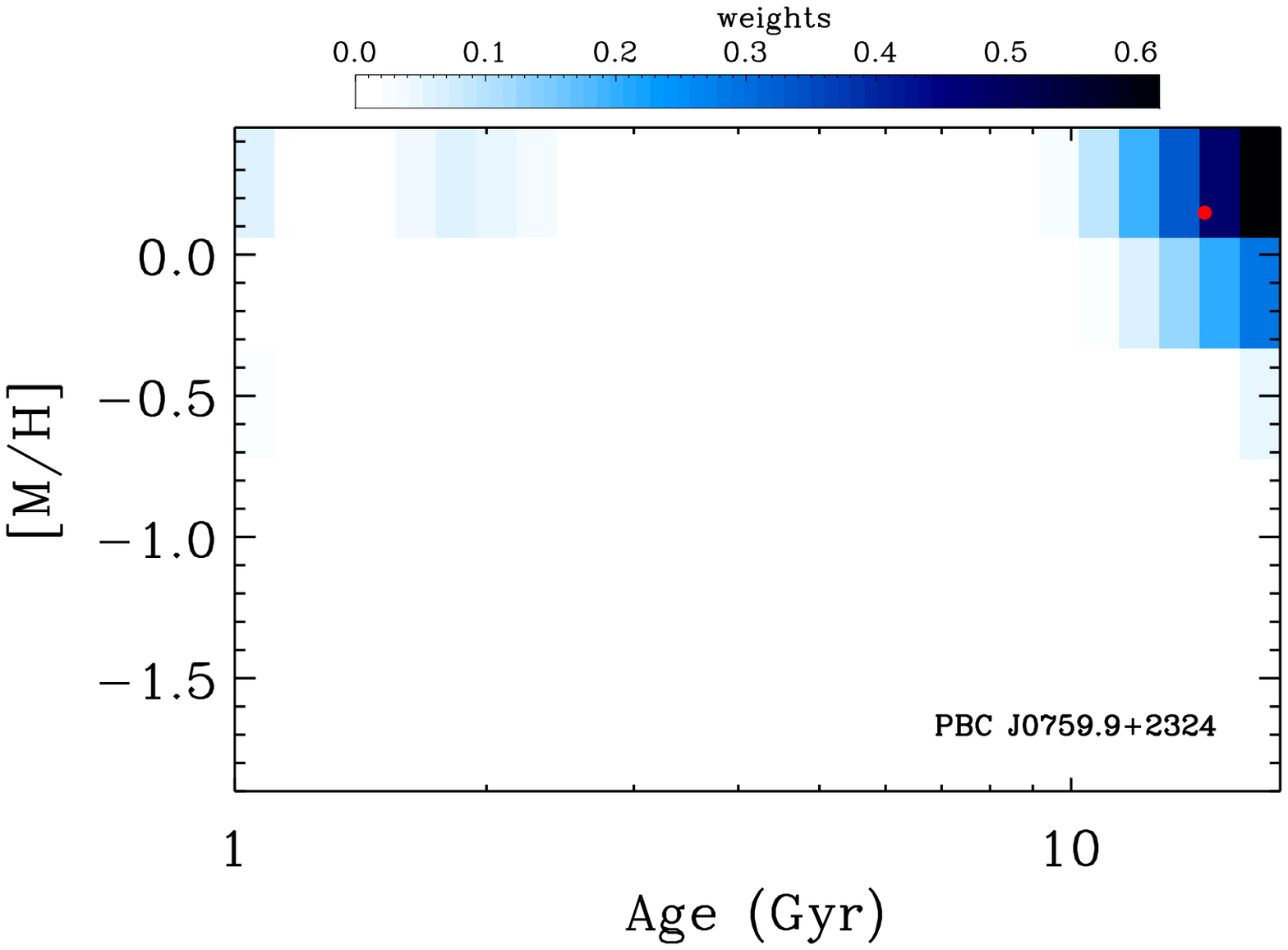}
              \includegraphics[angle=0.0,width=0.450\textwidth]{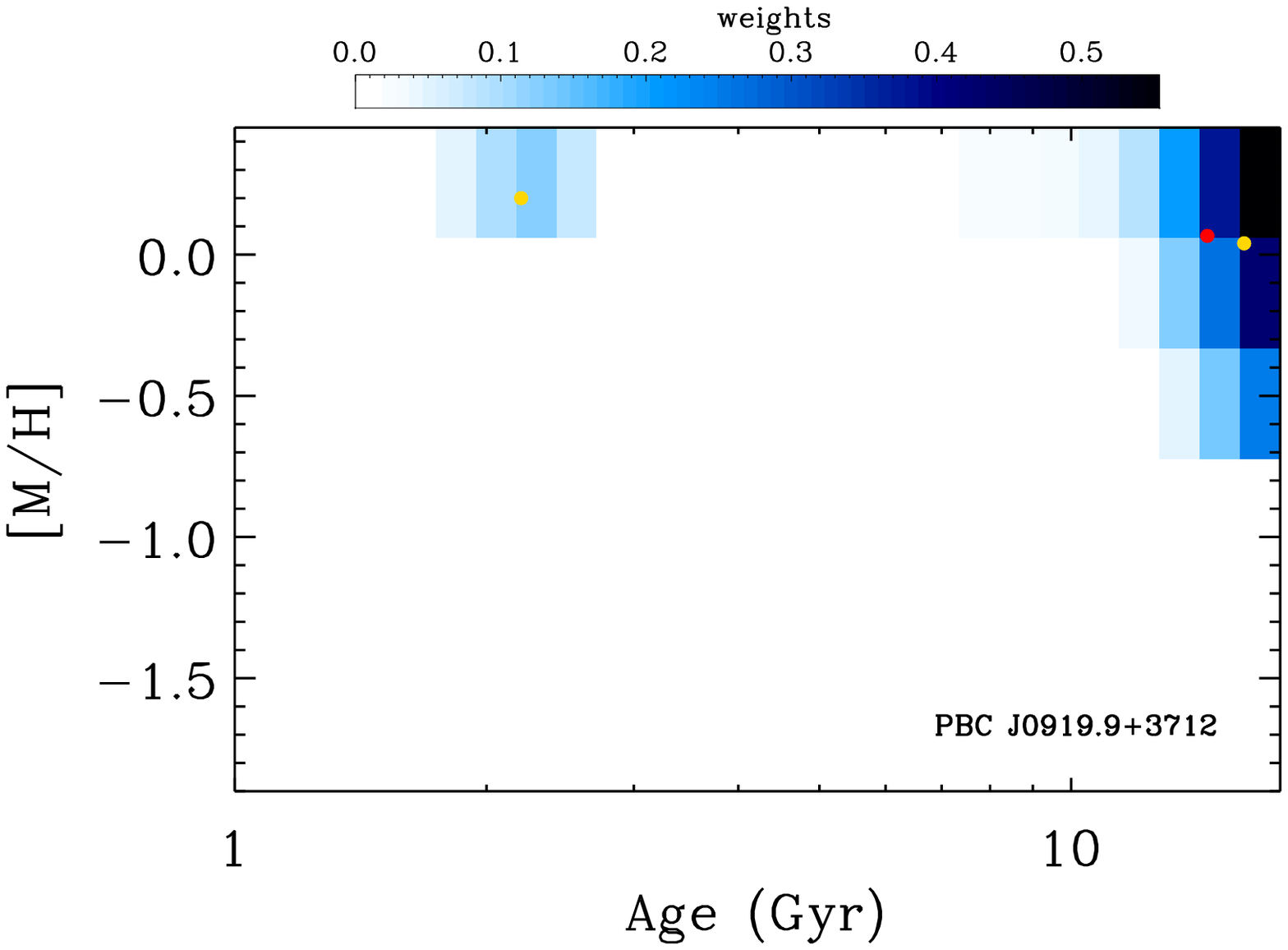}
           \caption{Age and metallicity obtained from the spectral
             fitting for IGR J01528-0326, IGR J02524-0829, IGR
             J04451-0445, IGR J18244-5622, IGR J18308+0928, PBC
             J0041.6+2534, PBC J0759.9+2324, PBC J0919.9+3712. The
             color scale refers to the mass fraction for each bin of
             age and metallicity. The red dot represents the
             mass-weighted age $\langle t/Gyr \rangle_M$ and the mass-
             weighted metallicity $\langle\,$\MH$\,\rangle_M$ for each
             galaxy. The yellow dots represent the mass-weighted age $\langle t/Gyr
                \rangle_M$ and the mass-weighted metallicity $\langle\,$\MH$\,\rangle_M$ in the case of two distinct stellar
                populations considered for the young and old component
                (see Sect. \ref{sec:risultati}).}
         \label{FigVibStab1}
\end{figure*}
%
\begin{figure*}
             \centering
              \includegraphics[angle=0.0,width=0.450\textwidth]{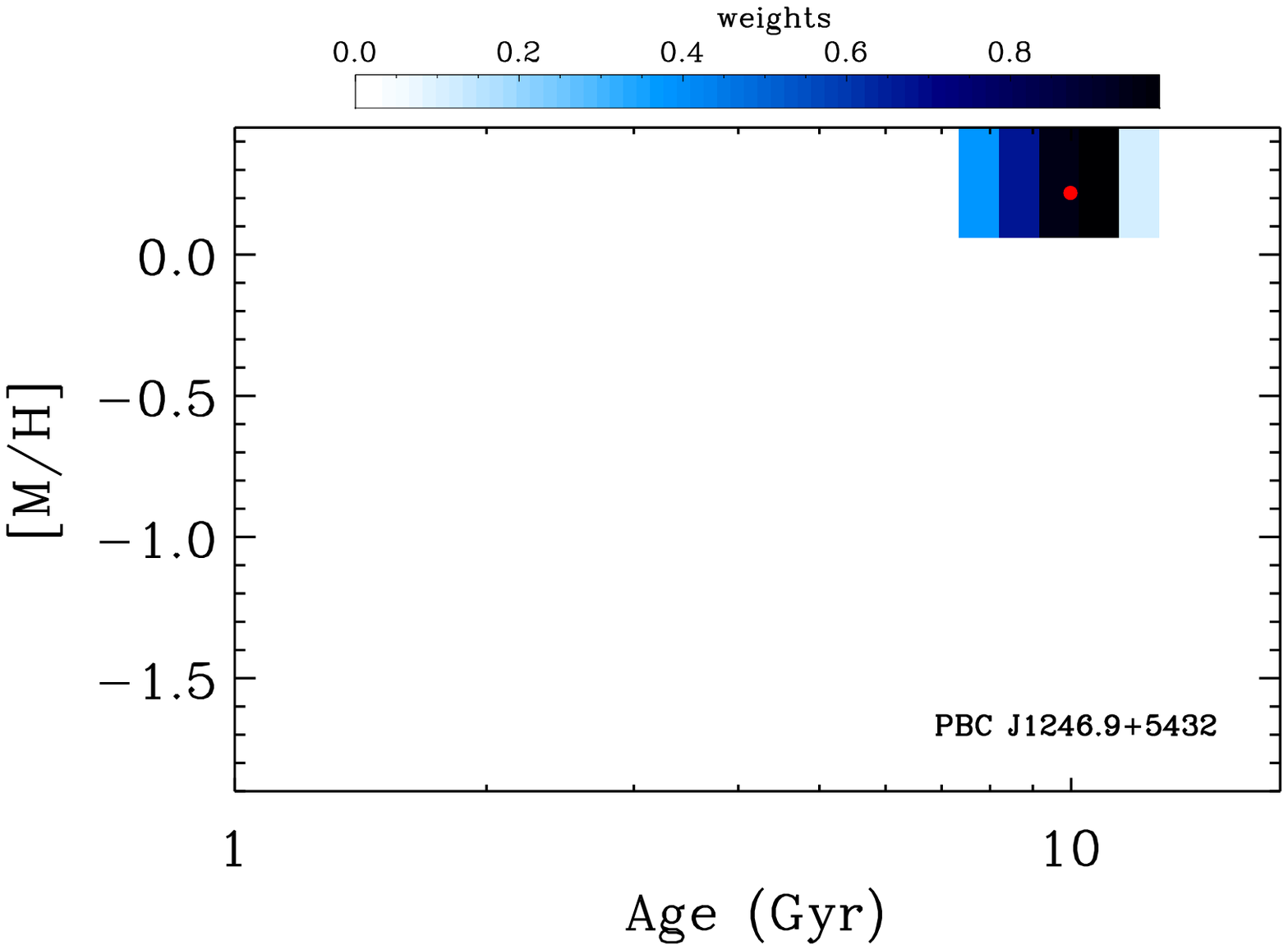}
              \includegraphics[angle=0.0,width=0.450\textwidth]{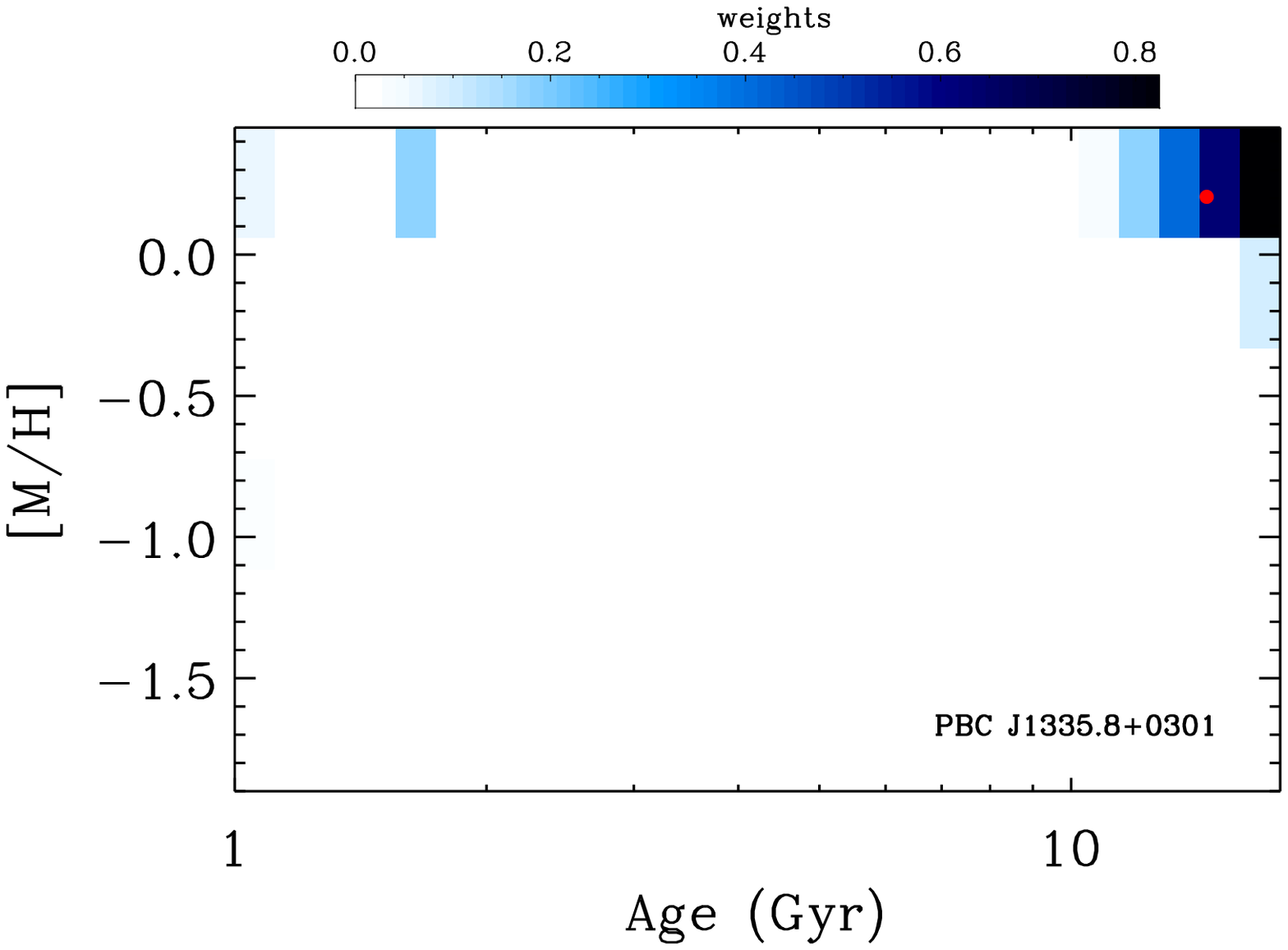}
              \centering
              \includegraphics[angle=0.0,width=0.450\textwidth]{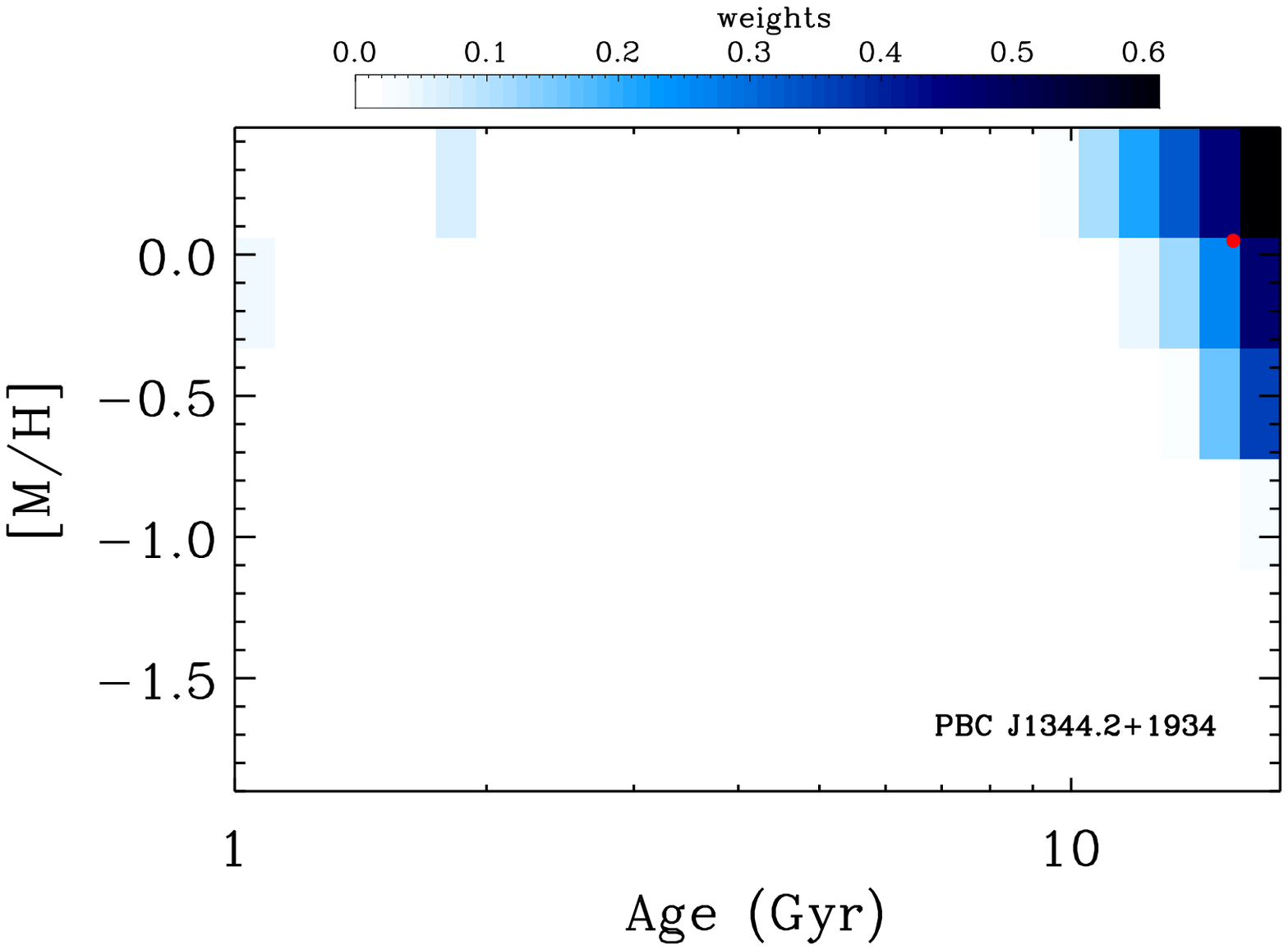}
              \includegraphics[angle=0.0,width=0.450\textwidth]{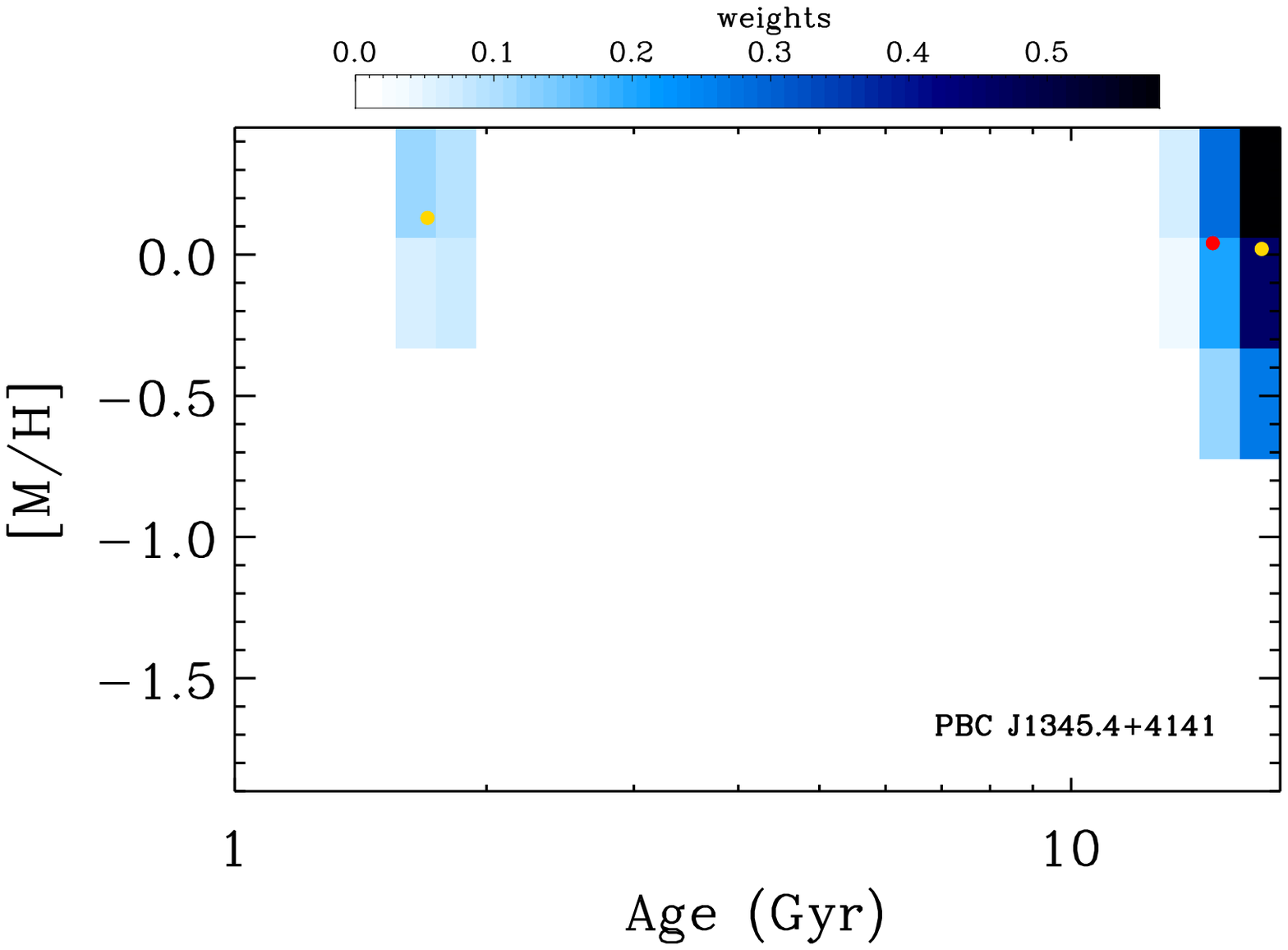}
             \centering
              \includegraphics[angle=0.0,width=0.450\textwidth]{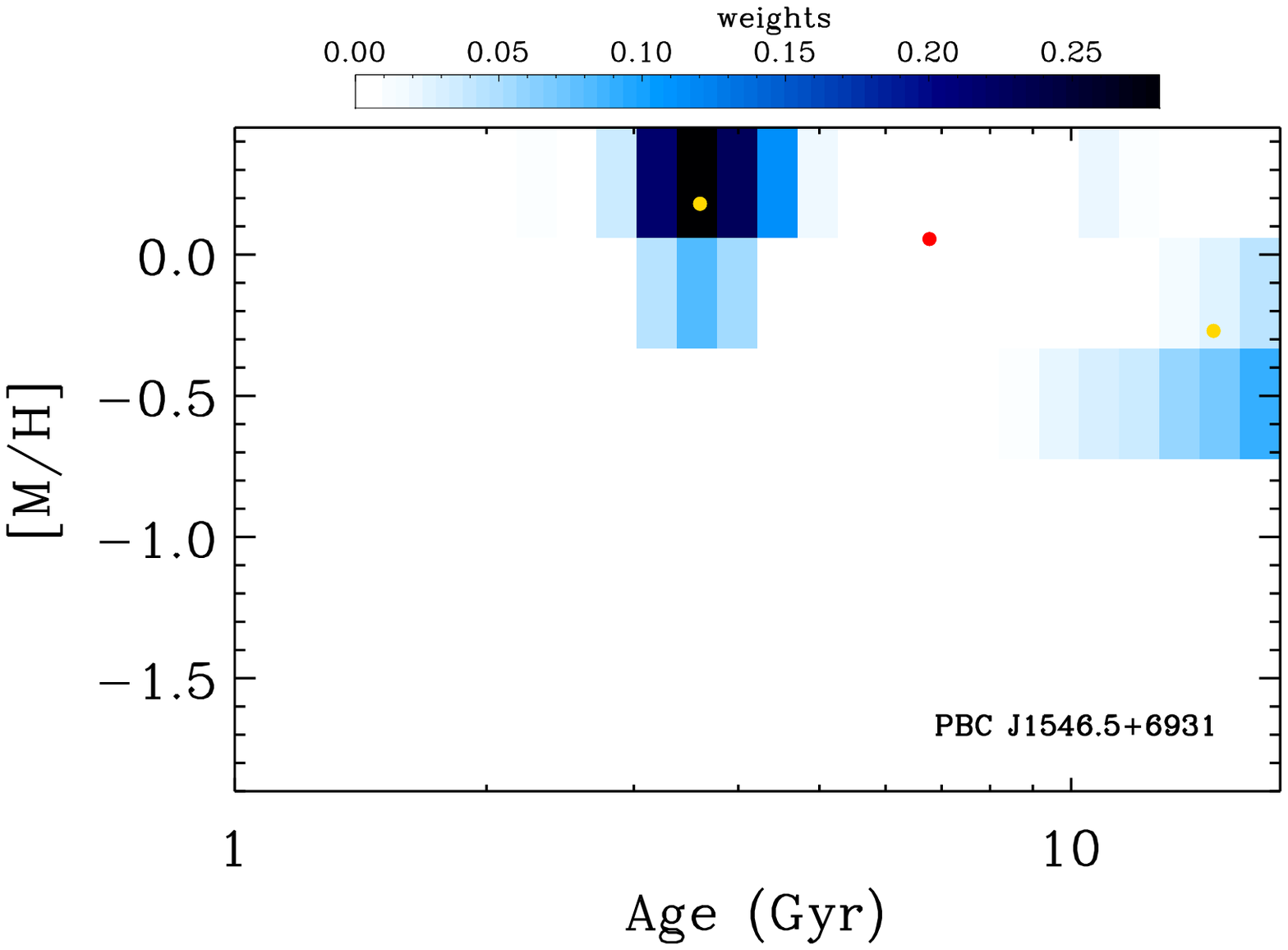}
              \includegraphics[angle=0.0,width=0.451\textwidth]{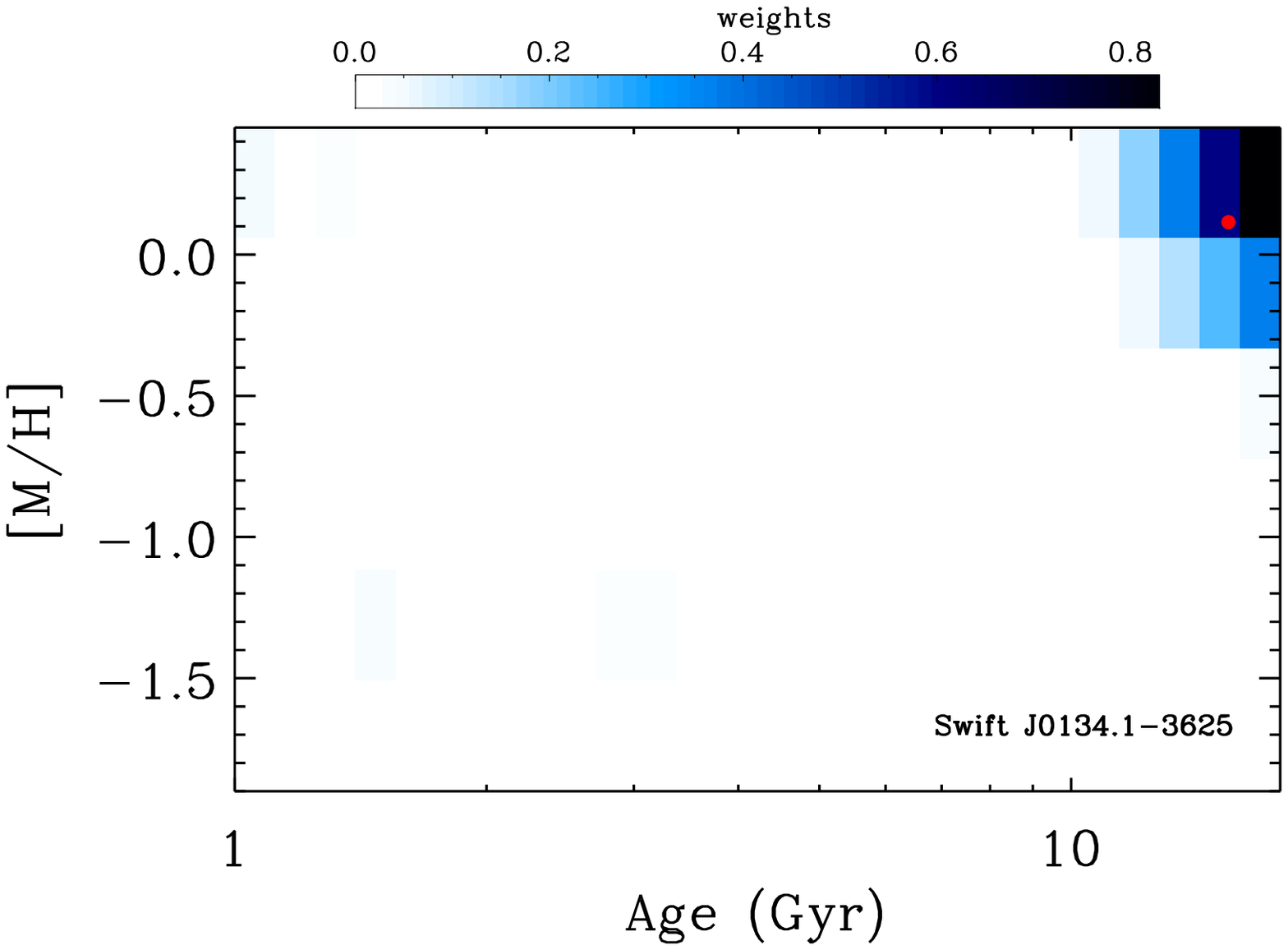}
            \centering
              \includegraphics[angle=0.0,width=0.451\textwidth]{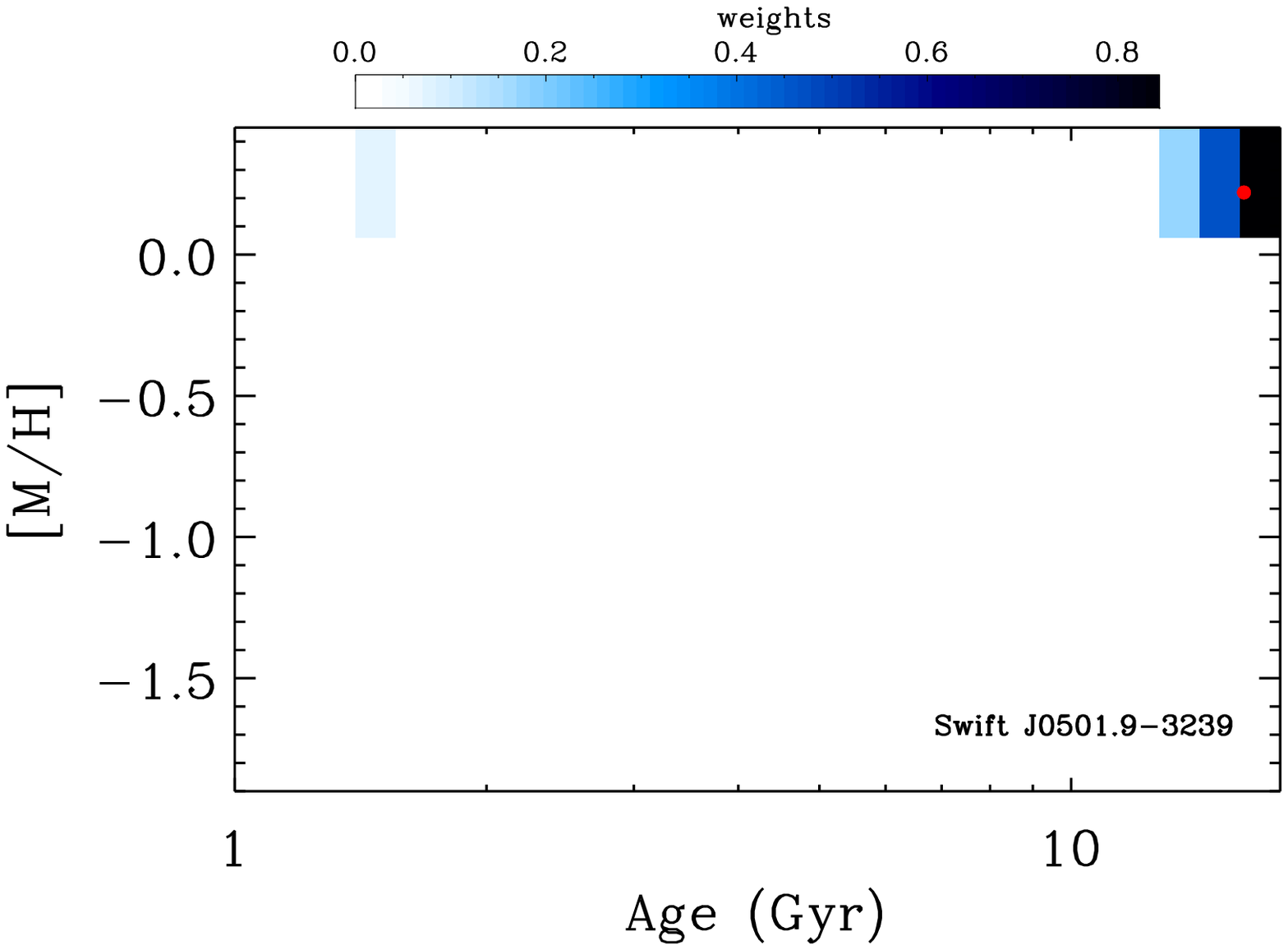} 
              \includegraphics[angle=0.0,width=0.450\textwidth]{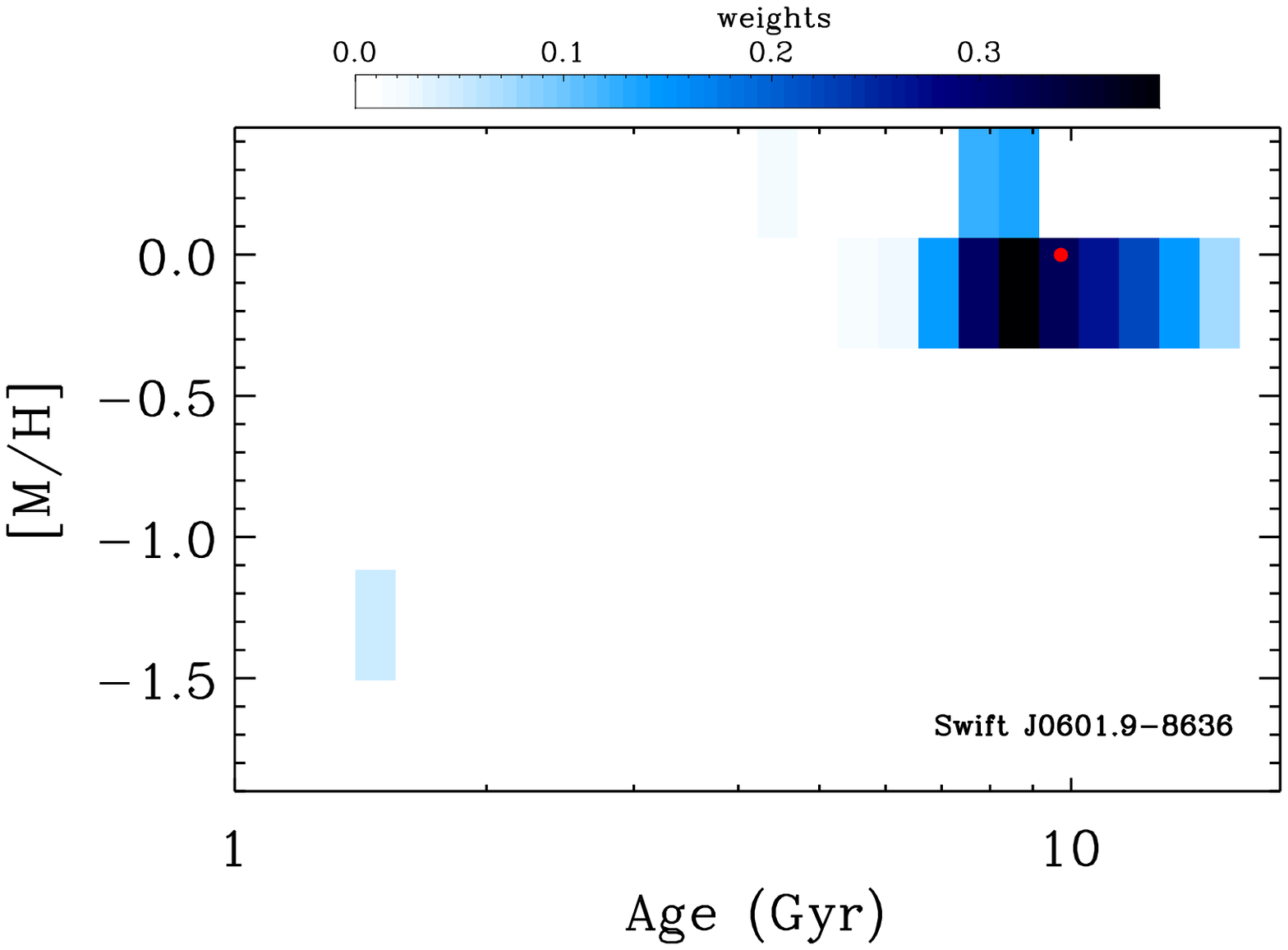}
              \caption{The same as in Figure \ref{FigVibStab1} for PBC
                J1246.9+5432, PBC J1335.8+0301, PBC J1344.2+1934, PBC
                J1345.4+4141, PBC J1546.5+6931, Swift J0134.1-3625,
                Swift J0501.9-3239, Swift J0601.9-8636. }
     \label{FigVibStab2}
\end{figure*}
%

\begin{figure*} 
              \centering
              \includegraphics[angle=0.0,width=0.450\textwidth]{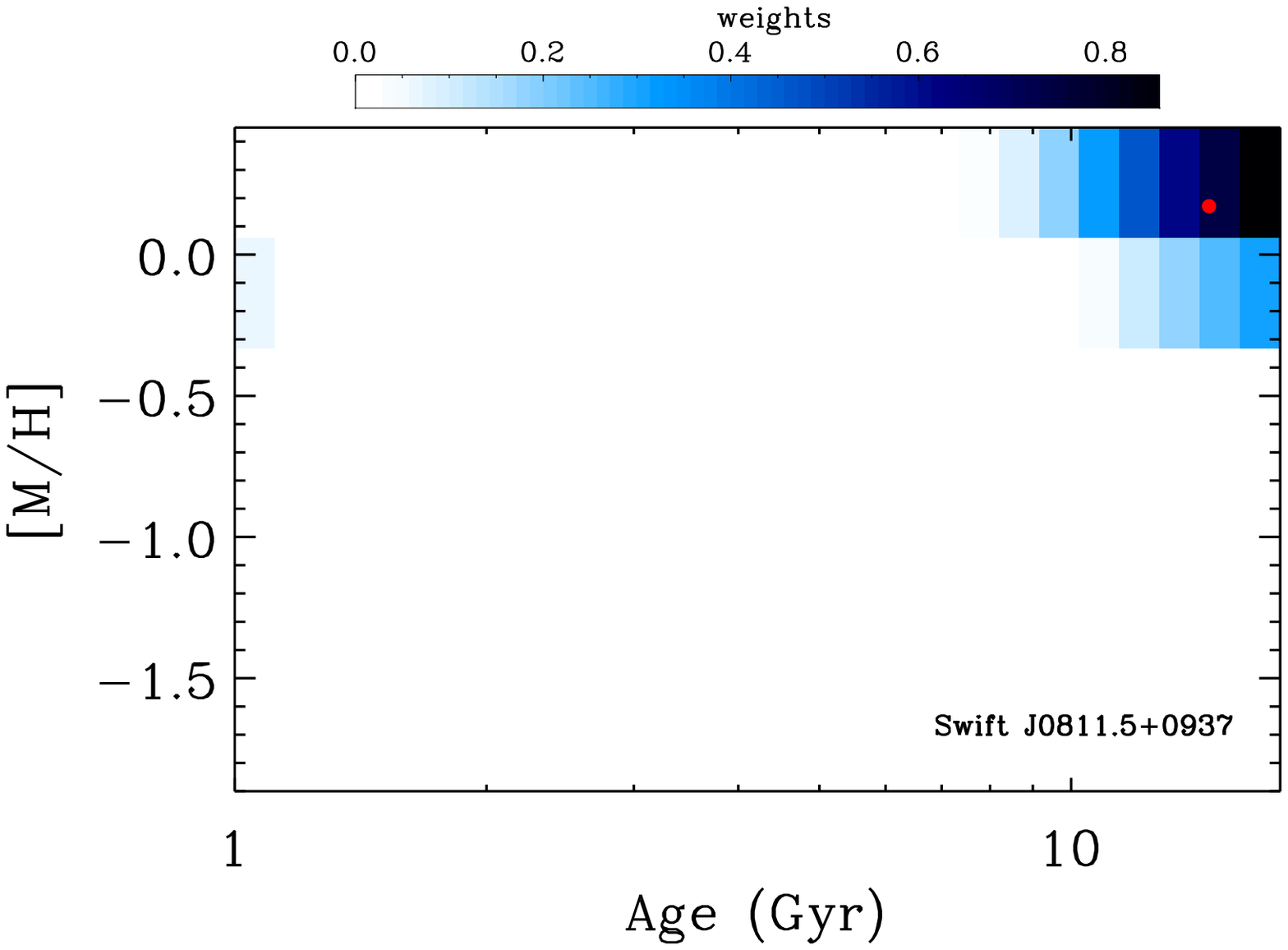}
              \includegraphics[angle=0.0,width=0.450\textwidth]{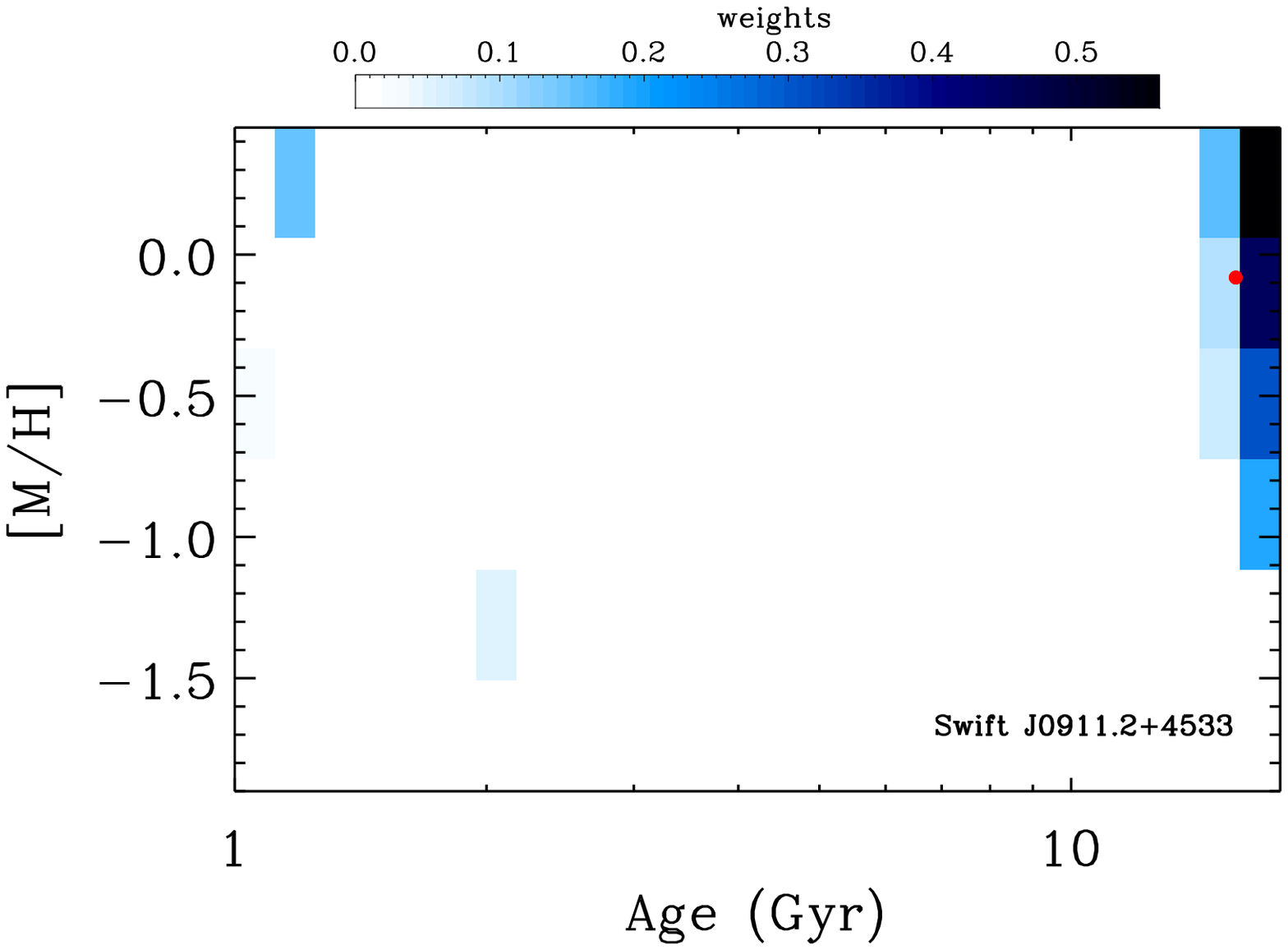}
              \centering
              \includegraphics[angle=0.0,width=0.450\textwidth]{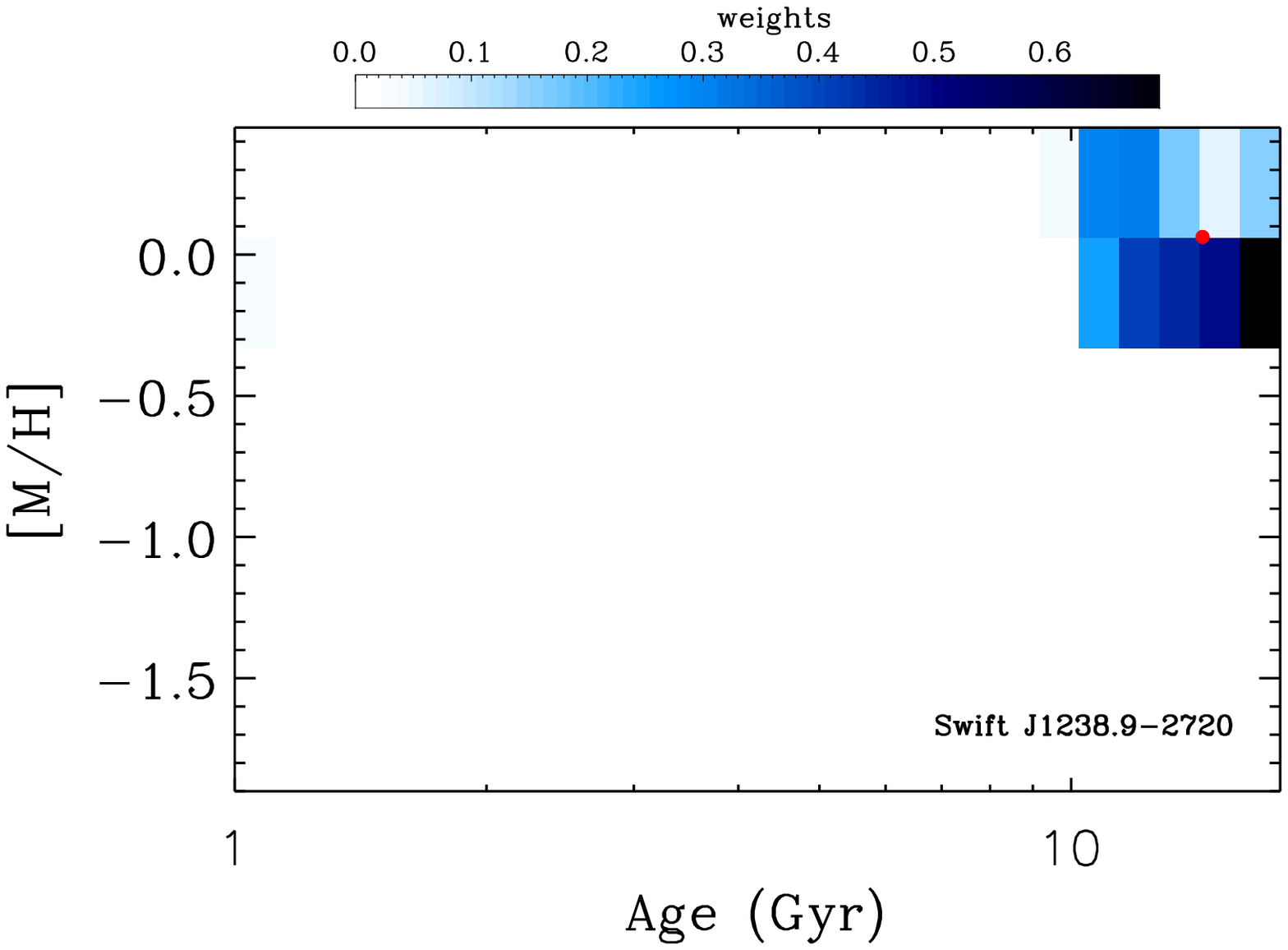}
             \caption{The same as in Figure \ref{FigVibStab1} for
                Swift J0811.5+0937, Swift
               J0911.2+4533, Swift J1238.9-2720. }
             \label{FigVibStab3}
\end{figure*}

The distribution of the age and metallicity templates used by the code
was, in general, smooth for most of the galaxies and indicated that
the bulk of stars in the considered region tends to be old.

To quantify this effect, for each galaxy we derived the mass-weighted
age $\langle t/Gyr \rangle_M$ and the mass-weighted metallicity
$\langle\,$\MH$\,\rangle_M$ of its stellar population. These values are
listed in Table \ref{Tab:results} and plotted in Figures
\ref{FigVibStab1}, \ref{FigVibStab2}, and \ref{FigVibStab3}.  The
color scale refers to the mass fraction in each bin of age and
metallicity. The errors on the age and metallicity given in Table
\ref{Tab:results} were obtained from photon statistics and CCD readout
noise, and they were calibrated through a series of Monte Carlo
simulations.

\begin{table}[t]
\caption{Mass-weighted age and metallicity measured for the galaxies
  in our sample.}
\label{Tab:results}

\begin{center}
\begin{tabular}{l r r}
\hline
\hline
\noalign{\smallskip}
\multicolumn{1}{c}{Galaxy} & 
\multicolumn{1}{c}{Age} & 
\multicolumn{1}{c}{Metallicity} \\
\noalign{\smallskip}
\multicolumn{1}{c}{name} & 
\multicolumn{1}{c}{(Gyr)} & 
\multicolumn{1}{c}{[M/H]} \\
\noalign{\smallskip}
\multicolumn{1}{c}{(1)} &
\multicolumn{1}{c}{(2)} &
\multicolumn{1}{c}{(3)} \\
\noalign{\smallskip}
\hline		    
\noalign{\smallskip}        	                		 
IGR J01528-0326    & 8.4$\pm$1.2  & 0.13$\pm$0.04\\ 
IGR J02524-0829    &15.2$\pm$0.8  & 0.01$\pm$0.02\\  
IGR J04451-0445    & 9.6$\pm$2.1  & 0.16$\pm$0.05\\ 
IGR J18244-5622    &12.8$\pm$1.1  &-0.17$\pm$0.04\\ 
IGR J18308+0928    &11.5$\pm$1.3  & 0.21$\pm$0.02\\  
PBC J0041.6+2534   & 5.0$\pm$1.1  & 0.19$\pm$0.03\\  
PBC J0759.9+2324   &14.4$\pm$0.7  & 0.15$\pm$0.01\\   
PBC J0919.9+3712   &14.5$\pm$0.6  & 0.06$\pm$0.02\\ 
PBC J0954.8+3724   &15.5$\pm$1.4  & 0.04$\pm$0.02\\  
PBC J1246.9+5432   &10.0$\pm$0.9  & 0.21$\pm$0.01\\ 
PBC J1335.8+0301   &14.5$\pm$0.7  & 0.20$\pm$0.02\\  
PBC J1344.2+1934   &15.6$\pm$1.0  & 0.04$\pm$0.04\\ 
PBC J1345.4+4141   &14.8$\pm$0.9  & 0.05$\pm$0.03\\  
PBC J1546.5+6931   & 6.8$\pm$1.7  & 0.05$\pm$0.03\\  
Swift J0134.1-3625 &15.4$\pm$1.3  & 0.11$\pm$0.03\\  
Swift J0501.9-3239 &15.2$\pm$1.4  & 0.10$\pm$0.02\\    
Swift J0601.9-8636 & 9.7$\pm$1.5  & 0.00$\pm$0.04\\  
Swift J0811.5+0937 &14.6$\pm$1.1  & 0.17$\pm$0.03\\ 
Swift J0911.2+4533 &15.7$\pm$0.9  &-0.08$\pm$0.02\\ 
Swift J1238.9-2720 &14.3$\pm$1.1  & 0.06$\pm$0.02\\ 
\noalign{\smallskip}
\hline				    	    			 
\end{tabular}
\end{center}
\begin{minipage}{8.5cm}
\begin{small}
NOTES: Col. (1): object name. Col. (2): mass-weighted age in Giga-years
  derived from the stellar population fitting. Col. (3): mass-weighted
  metallicity derived from the stellar population fitting.
\end{small}
\end{minipage}
\end{table}

In detail, almost all galaxies are characterized by an old
stellar population, with ages ranging from 8.4 Gyr to 15.7 Gyr.  There
are only 2 galaxies (PBC~J1546.5+6931 and PBC~J0041.6+2534) with $\langle
t/Gyr \rangle_M < 7$.  However, PBC~J1546.5+6931 shows a clear
bimodality in the distribution and, for this reason, $\langle t/Gyr
\rangle_M$ and $\langle\,$\MH$\,\rangle_M$ are not properly describing its
stellar population. Therefore, for this object we independently
derived the typical parameters of the two distinct stellar
populations. The younger one is characterized by $\langle t/Gyr
\rangle_M=3.6$ and $\langle\,$\MH$\,\rangle_M=0.18$ and it contributes to
$\sim70$\% of the total mass of the galaxy. The remaining $\sim30$\%
of the galaxy mass is due to a older population ($\langle t/Gyr
\rangle_M=14.8$) with subsolar metallicity ($\langle\,$\MH$\,\rangle_M=-0.27$).

Even though this is the most representative example of bimodality in
the stellar population distribution, there are other two galaxies,
namely PBC~J0919.9+3712 and PBC~J1345.4+4141, showing a second stellar
population accounting for more than 10\% of the total galaxy mass. The
final results for the three galaxies with two separate stellar
populations are listed in Table \ref{Tab:results2comp}. As can
  be noted from Table \ref{Tab:results2comp}, for both
  PBC~J0919.9+3712 and PBC~J1345.4+4141 the less massive component is
  a young one, which accounts for 12\% and 15\% of the total mass,
  respectively. The linear scales for these two objects are the
smallest in our sample ($\sim200$ pc) and this could have increased
our sensibility in detecting a young component existing in the
  nuclear region of the galaxy.
   \begin{figure}
              \includegraphics[angle=0.0,width=0.441\textwidth]{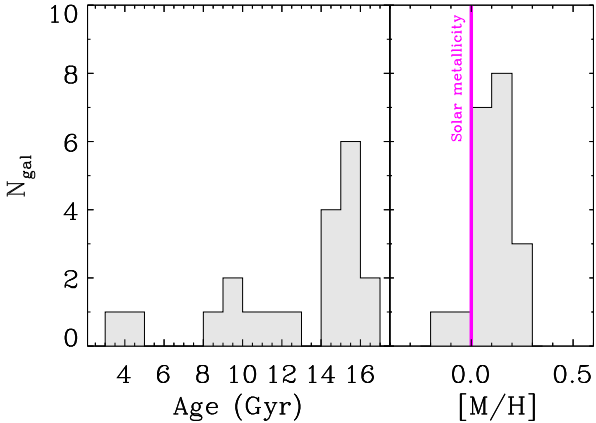}
    \caption[]{Distribution of mass-weighted ages (left panel) and
      metallicities (right panel) for the dominant stellar component in the
      sample of galaxies. The magenta vertical line indicates the
      value of the solar metallicity.}
    \label{fig:histograms}
   \end{figure}

It is interesting to note that an additional young component was also
found by \citet{cidfernandes2004} in $\sim30\%$ of the galaxies in
their sample. Their percentage, higher than what we found, could be
ascribed to the fact that we were observing a larger region of the
galaxy compared to what they did. For this reason we were less sensitive
to the young stellar populations which is, possibly, increasing
towards the nucleus of the galaxy \citep{cidfernandes2005III}.
  This could be the reason for the young component detected in
  PBC~J0919.9+3712 and PBC~J1345.4+4141. On the other hand, this is
not the case of PBC~J1546.5+6931 which is showing a dominant young
population and whose linear scale is bigger ($\sim2$ kpc). An
explanation for the global young stellar population of this galaxy
must be searched in its formation and evolutionary history.

Finally, the old ages we derived for the majority of the galaxies in
the sample are in agreement with the results obtained for a large
sample of infrared selected AGNs by \citet{chen09}. Those objects,
spanning different spectral classes and luminosities, all show clearly
the old stellar population dominating the total mass and no relevant
contributions from the young one.  These results are indirectly
confirming what \citet{schawinski2007} stated, i.e.  that Sy2s reside
very close to, or even lie in, the red sequence of galaxies.

The values of the mass-weighted age and metallicity that we obtained
considering only one stellar population (i.e. the one dominating the
mass) are shown in Figure \ref{fig:histograms} in the left and right
panel, respectively.

\begin{table}[t]
\caption{Mass-weighted age and metallicity measured for the young and
  old stellar populations considered separately. }
\label{Tab:results2comp}
\begin{center}
\begin{tabular}{c c c c c c}
\hline
\hline
\noalign{\smallskip}
\multicolumn{1}{c}{Galaxy} & 
\multicolumn{2}{c}{Age} & 
\multicolumn{2}{c}{Metallicity} &
\multicolumn{1}{c}{Mass } \\
\noalign{\smallskip}
\multicolumn{1}{c}{name} & 
\multicolumn{1}{c}{Young} &
\multicolumn{1}{c}{Old} &
\multicolumn{1}{c}{Young} &
\multicolumn{1}{c}{Old} &
\multicolumn{1}{c}{Old} \\
\noalign{\smallskip}
\multicolumn{1}{c}{} & 
\multicolumn{2}{c}{(Gyr)} & 
\multicolumn{2}{c}{[M/H]}&
\multicolumn{1}{c}{\%} \\
\noalign{\smallskip}
\hline		    
\noalign{\smallskip}        	                		 
PBC~J0919.9+3712&   2.2 & 16.1 &0.22 &0.04  & 88 \\ 
PBC~J1345.4+4141&   1.7 & 16.8 &0.13 &0.02  & 85 \\  
PBC~J1546.5+6931&   3.6 & 14.8 &0.18 &-0.27 & 30 \\  
\noalign{\smallskip}
\hline				    	    			 
\end{tabular}
\end{center}
NOTES: Col. (1): object name. Col. (2): mass-weighted age of the young and old stellar component. Col. (3): mass-weighted metallicity of the young and old stellar component. Col. (4): percentage of the total mass ascribed to the old stellar population.
\end{table}

The large majority of the sample galaxies are characterized by a
slightly supersolar mass-weighted metallicity (Figure
\ref{fig:histograms}, right panel).  The number distribution has a
median value of \MH$\,=0.08$ and spreads from super (\MH$\,=0.2$) to
sub-solar values (\MH$\,=-0.2$). However, it should be noted that
the fit used several different metallicities for the SSP, suggesting
the possible existence of more metallicity components.  Even when
considering two distinct stellar populations for PBC~J0919.9+3712,
PBC~J1345.4+4141, and PBC~J1546.5+6931 the values of the metallicity
remain inside this range (Table \ref{Tab:results2comp}). Once again,
this is consistent with the analysis done by \citet{chen09} and \citet
{lamura12} who claimed increasing metallicity going from starburst
towards Seyferts and LINERs.

   \begin{figure}
              \includegraphics[angle=0.0,width=0.481\textwidth]{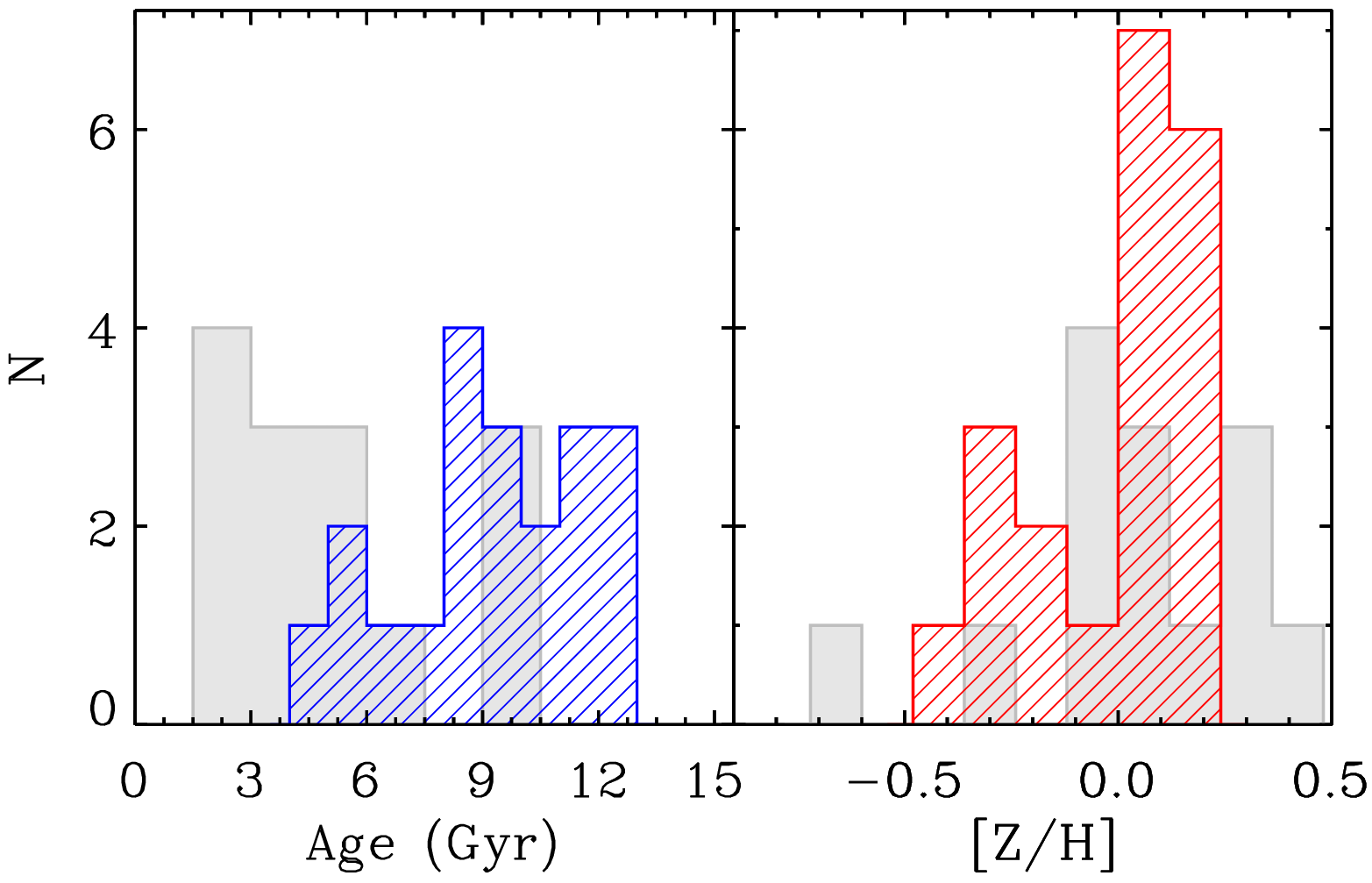}
    \caption[]{ The shaded histograms show the distribution of
      luminosity-weighted ages (left panel) and metallicities (right
      panel) for our sample of galaxies compared to the corresponding
      values derived by \citet{moreetal08} for a similar sample of
      spirals (grey solid histogram). }
    \label{fig:histogramscomp}
   \end{figure}

As a further step, we converted the mass-weighted ages and
metallicities to the corresponding luminosity-weighted values. To
perform the conversion, we adopted the $M/L$ ratios tabulated for the
SDSS $g$ filter by \cite{maraston05}.  As we expected, the
luminosity-weighted ages are slightly younger than the mass-weighted
ones, but they remain, with few exceptions, globally old spanning the
range between 6 and 12 Gyr. In Figure \ref{fig:histogramscomp} we plotted the
histogram of the luminosity-weighted ages compared to the one derived
for the high surface brightness (HSB) sample in \citet{moreetal08},
characterized by a similar morphological type distribution.

Even though the number of galaxies in our sample does not allow us to
trace a firm statistical conclusion, it is interesting to note that
the bulges in our sample are globally older than those hosted in
normal spirals. The metallicity of the bulges of LSB discs spans a
large range of values from high (\ZH$\,=\,0.30$ dex) to sub-solar
(\ZH$\,=\,-0.2$ dex) with a peak around slightly super-solar values
(Figure \ref{fig:histogramscomp}, right panel).  The distribution of
the metallicity in the galaxy sample is, instead, similar to the
one derived for the bulges of HSB galaxies
\citep{moreetal08}.

We performed an additional analysis looking for a possible correlation
between the mass-weighted ages and metallicities and the morphological
type of our galaxies.  \citet{cidfernandes2004} did not find any
relevant correlation between the host morphology and the stellar
population in the nuclear region of Sy2s, but in the case of
  non active galaxies the situation is less clear. Studying a sample
of spiral galaxies, \citet{thda06} and \citet{moreetal12} did not
observe any correlation between the age and metallicity of the stellar
population in the central region of the bulge and galaxy morphology,
whereas \citet{gandetal07} and \citet{moreetal08} found a mild
correlation, with the early-type galaxies ($T\,<\,0$) being older and
more metal rich than spirals ($T\,\geq\,0$).  The galaxies in our
sample span a large range of values of T type ($-2\,\leq\,T\,\leq\,5$)
homogeneously distributed without any decreasing trend in number going
from the early to the late type as observed in the
\citet{storchibergmann01} sample.

\begin{figure}
  \includegraphics[angle=90.0,width=0.491\textwidth]{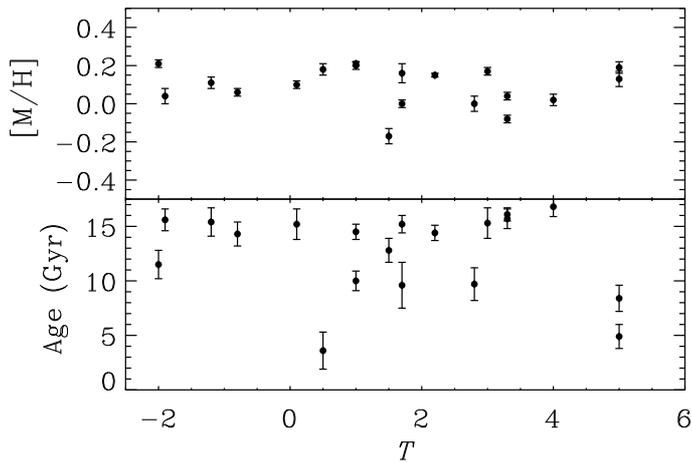}
  \caption[]{Correlation between the morphological type and the
    stellar population properties.  Top panel: the values for the
    mass-weighted metallicities are plotted as a function of the
    morphological type $T$. Bottom panel: as in  top panel for the
    mass-weighted ages in Gyr.}
  \label{fig:morph}
\end{figure}

We did not observe any relevant trend between the galaxy morphological
type and the age or metallicity of our galaxies (Figure
\ref{fig:morph}).  It is tempting to say that the AGN feedback is, in
some way, acting homogenizing the stellar populations in the central
(few kpc) region of galaxies with different morphological
type. However, the low number of galaxies with $T\leq\,0$ and the
shallow relation between morphology and stellar populations obtained
even considering non active galaxies prevented us from claiming any
strong conclusion on this aspect.

\section{Summary and Conclusions}
\label{sec:conclusions}

In the last years our collaboration unveiled the nature of more than 250
X-ray emitting sources, 158 of which were identified as AGNs. In this
paper we presented the stellar population analysis of the host galaxy
which was performed on a carefully selected sample of 20 objects with
the goal to contribute to understand the still debated connection
between the central AGN and the properties of the host galaxy, as well as the
effect of the AGN feedback. The detailed analysis of the stellar
population can give further important constraints on this topic.

The area on which we were measuring the galaxy properties ranges from
200 pc to 2 kpc allowing us to measure the stellar populations of the
bulge close to the center of the galaxy where the effects of the AGN
feedback on the host are expected to be relevant.

The spectral fitting method, based on PPXF and GANDALF, was applied to
the galaxy spectra to measure the stellar populations properties of
the sample. In particular, we obtained for each galaxy a mass-weighted
age and a mass-weighted metallicity.  The values of the mass-weighted
metallicity span the range $-0.2\leq\,$\MH$\,\leq0.2$ with a median value
of \MH$\,=\,0.08$. The large majority of our objects (19 galaxies out of
20, i.e. 95\%) show an old stellar population, with ages older than 8
Gyrs.  Three of them are characterized by a bimodal distribution with
a non negligible contribution from young stars. In detail, we found
that PBC~J1546.5+6931 is dominated by a young stellar population
accounting for $\sim70$\% of the total mass, while, in the case of
PBC~J0919.9+3712 and PBC~J1345.4+4141, the young stellar component is
less massive than the old one. For the former galaxy the nature of the
young stellar component is probably related to the formation and
evolution of the galaxy itself.  Regarding the latter galaxies there
would be a possible starforming activity in their central region.

Even investigating the luminosity-weighted ages for the galaxies in
our sample the old nature of their stellar population is
confirmed. The comparison of their ages with those obtained for a
similar sample in terms of morphological type showed that bulges
hosted in AGN are globally older than those hosted in non active
counterparts. However, it should be noted that, as expected, the
contribution of the young component to the total light of the galaxy
is greater than the one to the total mass.

Our results suggest that AGN feedback acts on the first kpc of the
galaxy, decreasing the efficiency of the star formation, trough
  different processes.

The combination of truncation and suppression \citep{schaetal09} could
be responsible for disrupting the gas starforming reservoir
\citep{davis12} in early times, when the gas is wiped out from the
strong AGN emission in the center
\citep{dimatteo2005,springel05,fontanot11}, and in the recent times,
the phase of the AGN regulates and quenches the residual star
formation of the host galaxy \citep{sturetal11} maintaining the global
red color and old stellar populations observed also in this work.

Radiative feedback and cooling flow of gas to the center could also,
as proposed by \citet{ciotti10} and \citet{novak11}, exhaust the gas in the few
central kpcs region of the galaxy and decrease the star
formation, with the final consequence that the bulk of the observed
stellar populations in this region is, with very few exceptions, old.

These scenarios are indirectly supported by the lack of relations
between the stellar population properties and the morphological ones,
in the sense that in the central region the existence of an AGN
influences the properties of the galaxy much more than the formation
and evolution, described by the morphological type, itself.

\begin{acknowledgements}
We thank Valentin Ivanov and Lodovico Coccato for  useful
discussions.  LM acknowledges financial support from Padua University
grant CPS0204. LM and VC acknowledge the Universidad Andr\'es Bello in
Santiago del Chile for hospitality while this paper was in progress.
PP is supported by the INTEGRAL ASI-INAF grant No. 033/1070.  NM
acknowledges the Pontificia Universidad Cat\'olica de Chile for
hospitality while this paper was in progress. DM and GG are supported
by the Basal CATA Center for Astrophysics and Associated Technologies
PFB-06. GG is supported by fondecyt grant 1120195.

\end{acknowledgements}


\bibliographystyle{aa} 
\bibliography{loreb.bib} 


\end{document}